\documentclass[10pt,conference]{IEEEtran}

\usepackage{cite}
\usepackage{lineno}
\usepackage{amsmath,amssymb,amsfonts}
\usepackage{algorithmic}
\usepackage{graphicx}
\usepackage{textcomp}
\usepackage{xcolor}
\usepackage{booktabs}
\usepackage{colortbl}
\usepackage{color}
\usepackage{xspace}
\usepackage{url}
\usepackage{balance}
\usepackage{tabulary}
\usepackage{array}
\usepackage{graphicx}
\usepackage[ruled,lined,linesnumbered,vlined,algo2e]{algorithm2e}
\usepackage{subfig}
\usepackage{listings}
\usepackage{lipsum}
\usepackage{hyperref}
\usepackage{graphicx}
\usepackage{amsmath,amssymb}
\usepackage[T1]{fontenc}
\usepackage{amsmath, amsthm, amssymb}
\usepackage[ansinew]{inputenc}
\usepackage[many]{tcolorbox}
\usepackage{color}
\usepackage{xcolor}
\usepackage{extarrows}
\usepackage{booktabs}
\usepackage{multirow}
\usepackage{colortbl}
\usepackage{tabularx}
\usepackage{ulem}
\usepackage{enumitem}
\usepackage{lipsum}
\graphicspath{{fig/}}
\usepackage[justification=centering]{caption}

\definecolor{Green}{RGB}{0,180,0}

\newcommand{\github}{\url{https://github.com/kbcao/sequer}}
\newcommand{\tool}{\texttt{SEQUER}}

\newcommand{\cameraReady}[1]{\textcolor{black}{#1}}

\newcommand{\tabincell}[2]{\begin{tabular}{@{}#1@{}}#2\end{tabular}}

\def\BibTeX{{\rm B\kern-.05em{\sc i\kern-.025em b}\kern-.08em
    T\kern-.1667em\lower.7ex\hbox{E}\kern-.125emX}}
\begin{document}

\title{Automated Query Reformulation for Efficient Search based on Query Logs From Stack Overflow}

\author{\IEEEauthorblockN{Kaibo Cao\IEEEauthorrefmark{2}, Chunyang Chen\IEEEauthorrefmark{3}\IEEEauthorrefmark{1}, Sebastian Baltes\IEEEauthorrefmark{4}, Christoph Treude\IEEEauthorrefmark{4}, Xiang Chen\IEEEauthorrefmark{5}\IEEEauthorrefmark{1}}
\IEEEauthorblockA{\IEEEauthorrefmark{2}\textit{Software Institute},
\textit{Nanjing University}, China}
\IEEEauthorblockA{\IEEEauthorrefmark{3}\textit{Faculty of Information Technology}, 
\textit{Monash University}, Australia}
\IEEEauthorblockA{\IEEEauthorrefmark{4}\textit{School of Computer Science}, 
\textit{University of Adelaide}, Australia}
\IEEEauthorblockA{\IEEEauthorrefmark{5}\textit{School of Information Science and Technology}, 
\textit{Nantong University}, China}
imkbcao@gmail.com,  chunyang.chen@monash.edu, \{sebastian.baltes, christoph.treude\}@adelaide.edu.au, xchencs@ntu.edu.cn
}

\maketitle

\begingroup
\renewcommand{\thefootnote}{}
\footnotetext[1]{\IEEEauthorrefmark{1} Corresponding Authors}
\endgroup

\begin{abstract}
    As a popular Q\&A site for programming, Stack Overflow is a treasure for developers.
    However, the \cameraReady{amount of} questions and answers on Stack Overflow make it difficult for developers \cameraReady{to efficiently locate the information they are looking for}.
    There are two gaps leading to poor search results: the gap between the user's intention and the textual query, and the semantic gap between the query and the post content.
    Therefore, developers have to constantly reformulate their queries by correcting misspelled words, \cameraReady{adding limitations to certain programming languages or platforms}, etc.
    As query reformulation is tedious for developers, especially for novices, we propose an automated software-specific query reformulation approach based on deep learning.
    With query logs provided by Stack Overflow, we construct a large-scale query reformulation corpus, including the original queries and corresponding reformulated ones.
    Our approach trains \cameraReady{a Transformer model} that can automatically generate candidate reformulated queries when given the user's original query.
    The evaluation results show that our approach outperforms five state-of-the-art baselines, and achieves a 5.6\% to 33.5\% boost in terms of $\mathit{ExactMatch}$ and a 4.8\% to 14.4\% boost in terms of $\mathit{GLEU}$.
\end{abstract}


\section{Introduction}
\label{sec:intro}

Stack Overflow is the most popular question and answer (Q\&A) site for programming-related knowledge sharing and acquisition. 
Over the past decade, Stack Overflow has accumulated a large amount of user-generated \cameraReady{content}, making it a valuable repository of software engineering knowledge. 
When developers encounter a specific programming question \cameraReady{such as \textit{how to use this library?}, \textit{what is the difference between two languages?}~\cite{huang2018tell}, or \textit{how to understand this concept?}}~\cite{chen2016mining,chen2016techland}, they tend to use Stack Overflow to find answers~\cite{abdalkareem2017developers}.
To assist developers in finding the knowledge \cameraReady{they are looking for} in such a large-scale knowledge repository, Stack Overflow provides a search engine\footnote{\url{https://stackoverflow.com/search}}, which supports free text search as well as an advanced search with metadata filters.

However, even using the provided search engine, it is still not easy for developers to effectively find what they want~\cite{xia2017developers,chen2016towards}.
There are two reasons for unsatisfactory search results.
First, there exists a certain semantic gap between the users' query intention and input queries. 
\cameraReady{This} means that it is difficult for users to accurately express their query intention with a few keywords ~\cite{datta2008image, zha2010visual}.
For example, a developer wants to search for the usage of nested lists. However, they may not know how to express this concept accurately and may use a term like ``\textit{list in list}'' as the query.
This query is imprecise, and this kind of semantic gap can pose a significant challenge to the search engine of Stack Overflow.
Second, a certain semantic gap exists between the users' queries and the \cameraReady{text} content in the relevant posts.
It means the same meaning may be described in different \cameraReady{ways} with few overlapping words.
For example, a developer may input the query ``\textit{sorting in linear time}''. However, the relevant post titled ``\textit{sorting with $O(n)$ complexity}'' cannot be retrieved by the search engine. 
Moreover, abbreviations, synonyms, or even misspelling~\cite{chen2017unsupervised, chen2019sethesaurus} can also lead to this kind of semantic gap.

\begin{figure}[htbp]
    \centering
    \vspace{-2mm}
    \includegraphics[width=0.4\textwidth]{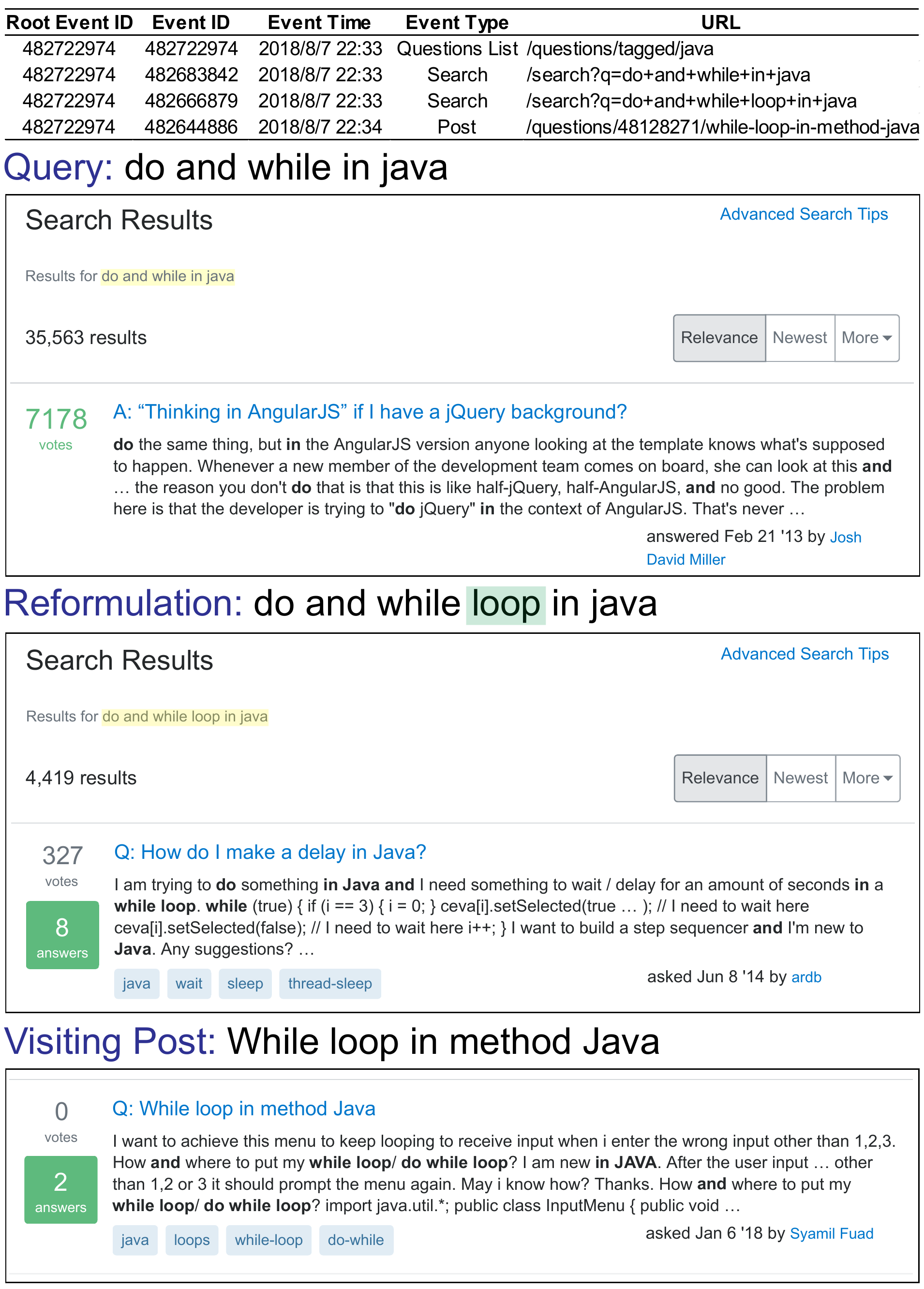} 
    \caption{An illustrative example of query reformulation in the users' activity logs}
    \vspace{-5mm}
    \label{fig:intro-datasetSample}
\end{figure}

To alleviate the above semantic gaps, developers may constantly reformulate their queries until the query reflects their real query intention and leads to relevant posts.
By analyzing users' query logs provided by Stack Exchange Inc., the company behind Stack Overflow, we find that 24.62\% of the queries are reformulated before visiting a post, and that developers reformulate their queries 1.46 times on average before clicking on a result.
Fig.~\ref{fig:intro-datasetSample} shows an illustrative example.
The user first performs the query ``\textit{do and while in java}'', which yields a large number of irrelevant results.
The reason is that the search engine uses \textit{or} to connect all the words in the query, and the search engine cannot distinguish ``\textit{do}'' and ``\textit{while}'' as domain-specific keywords.
Then, the user adds the word ``\textit{loop}'' to this query to make it more explicit, which leads to a \cameraReady{potentially} \cameraReady{desired} post.
The process of modifying a given query to find a satisfactory \cameraReady{search result}~\cite{jansen2009patterns} is called query reformulation.
Understanding query reformulation~\cite{sloan2015term, bing2015web, jiang2014learning, rieh2006analysis, huang2009analyzing} has become an important issue in designing effective information retrieval systems.

To investigate how developers reformulate their queries, we first perform a formative study on the users' activity logs from Stack Overflow.
Based on the logs from 2,216,219 users between December 2017 and November 2018, we extract 4,631,756 queries from 3,125,427 sessions.
After analyzing these queries, we find that query reformulation has \cameraReady{certain} common patterns (in Section~\ref{sec:emp-whyreformulate}).
For example, \cameraReady{users} may fix misspellings, such as revising ``\textit{serive}'' to ``\textit{service}''; 
they may generalize their queries to expand the scope, such as revising ``\textit{open calendar react native}'' to ``\textit{open other app react native}'';
they may add constraints of programming languages or platforms to the query, such as revising ``\textit{read file}'' to ``\textit{java read file}'';
they may remove information that is too detailed for retrieving relevant posts, such as revising ``\textit{Unable to import module 'copy': /var/task/psycopg2/\_psycopg.so: ELF file's phentsize not the expected size}'' to ``\textit{ELF file's phentsize not the expected size}''.

As the query reformulation process is tedious \cameraReady{for} developers, especially for novices, we propose a \textbf{S}oftware-sp\textbf{E}cific \textbf{QUE}ry \textbf{R}eformulation approach ({\tool}) based on deep learning to help automatically reformulate their queries.
Based on large-scale query logs from Stack Overflow \cameraReady{provided} under a non-disclosure agreement, we first extract query reformulation pairs \cameraReady{consisting of original and reformulated queries} and then
adopt an attention-based Transformer to automatically learn the query reformulation patterns based on the extracted query reformulation pairs.
Given the original query, the trained model can suggest a list of candidate reformulated queries.
We evaluate the quality of the reformulated queries of our approach with large-scale archival manual reformulation results.
The evaluation results show that, in terms of $\mathit{ExactMatch@10}$ and $\mathit{GLEU}$, our approach achieves 12.48\% and 7.79\% improvement on average compared with the sequence model based baselines (i.e., seq2seq with attention~\cite{sutskever2014sequence} and HRED-qs~\cite{sordoni2015hierarchical}), 
achieves 30.79\% and 6.61\% improvement compared with Google Prediction Service~\cite{cornea2014providing}, 
and achieves 33.5\% and 14.41\% improvement compared with grammatical error correction tools.

In summary, we make the following contributions:
\begin{itemize}
	\item We distill unique insight into developers' query reformulation patterns based on large-scale real-world query logs from Stack Overflow. 
	
	\item According to the insights from our empirical study, we propose a software-specific query reformulation approach {\tool} based on an attention-based Transformer. 
	
	\item We evaluate the quality of the reformulated queries generated by our approach {\tool} with large-scale archival manual reformulation results. 
	
	\item We implement a browser plugin\footnote{\github} for supporting automated software-specific query reformulation in practice. 
\end{itemize}

\section{Data Collection}
\label{sec:data}

\cameraReady{The dataset we used is based on a larger} dataset containing all internal HTTP requests processed by Stack Overflow's web servers within one year (747,421,780 requests from December 2017 to November 2018).
Internal means that the dataset only contains requests with a referrer URL from \texttt{stackoverflow.com}.
If a user, for example, reached a Stack Overflow post by clicking on a Google search result and then triggered a search within Stack Overflow, only the second (internal) search request would be included in the dataset, not the request for the post having a Google-related referrer.
For each HTTP request, the dataset provides an anonymized user identifier that represents logged-in registered users as well as users identified by a browser cookie or users identified by their IP address.
\cameraReady{This dataset also assigns certain event types to the requests (e.g., searching, post visiting, or question list browsing), depending on their target URL.}

Before extracting the event sequences relevant \cameraReady{to} this study, we \cameraReady{preprocess the data as follows}.
First, we group all events per user identifier and then order them chronologically.
Second, we utilize heuristics based on the timestamps and request targets to filter out bot traffic \cameraReady{and} event sequences merely consisting of page refreshes.
Third, to distinguish between individual sessions, we group the events into \cameraReady{sequences of events that are not more than six minutes apart}, following Sadowski et al.'s approach~\cite{sadowski2015developers}.
Finally, we \cameraReady{add} an additional filtering step to avoid gaps in the data \cameraReady{caused by} the focus on internal requests.
A user may, for example, follow external links in Stack Overflow posts and then navigate back to Stack Overflow or open multiple posts in parallel browser tabs.
For our analysis of query reformulation on Stack Overflow, we focus on \cameraReady{complete} linear navigation sequence, that \cameraReady{is} sequence where the referrer of one request matches the target URL of the previous request. 

After the above data preprocessing, we get a dataset of \cameraReady{complete} linear navigation sequences.
As shown in the table in Fig.~\ref{fig:intro-datasetSample}, each event is represented by one row with five attributes: \textit{RootEventId}, \textit{EventId}, \textit{EventTime}, \textit{EventType}, and \textit{URL}.
Specifically,
the \textit{RootEventId} refers to the \textit{EventId} of the first event in the session,
the \textit{EventId} is a unique ID that can identify an event,
the \textit{EventTime} indicates the UTC time when the event occurred,
the \textit{EventType} shows the type of the event (possible values of event type are $\{$\textit{Search}$, $\textit{Post}$, $\textit{QuestionsList}$, $\textit{Home}$, $\textit{Tags}$, $\textit{PostHistory}$\}$),
the \textit{URL} is the web request that triggers this event, which contains, for example, \cameraReady{the query content or the post ID, depending on the \textit{EventType}}.

Our dataset contains 42,173,522 events from 16,164,506 sessions generated by 9,712,878 users, in which 9,046,179 events are queries.
On Stack Overflow, the post visit event and the query event are the two most common operations performed by the users, accounting for 46.21\% and 21.45\% \cameraReady{of the events} respectively.
Users often \cameraReady{reach} a post in three ways: links in the post (58.75\%), search (17.25\%), and question list (5.69\%).
That is, in addition to the navigation between posts, search is the most common way for the users to find a post.

\begin{figure}[htbp]
    \centering
    \vspace{-2mm}
    \includegraphics[width=0.4\textwidth]{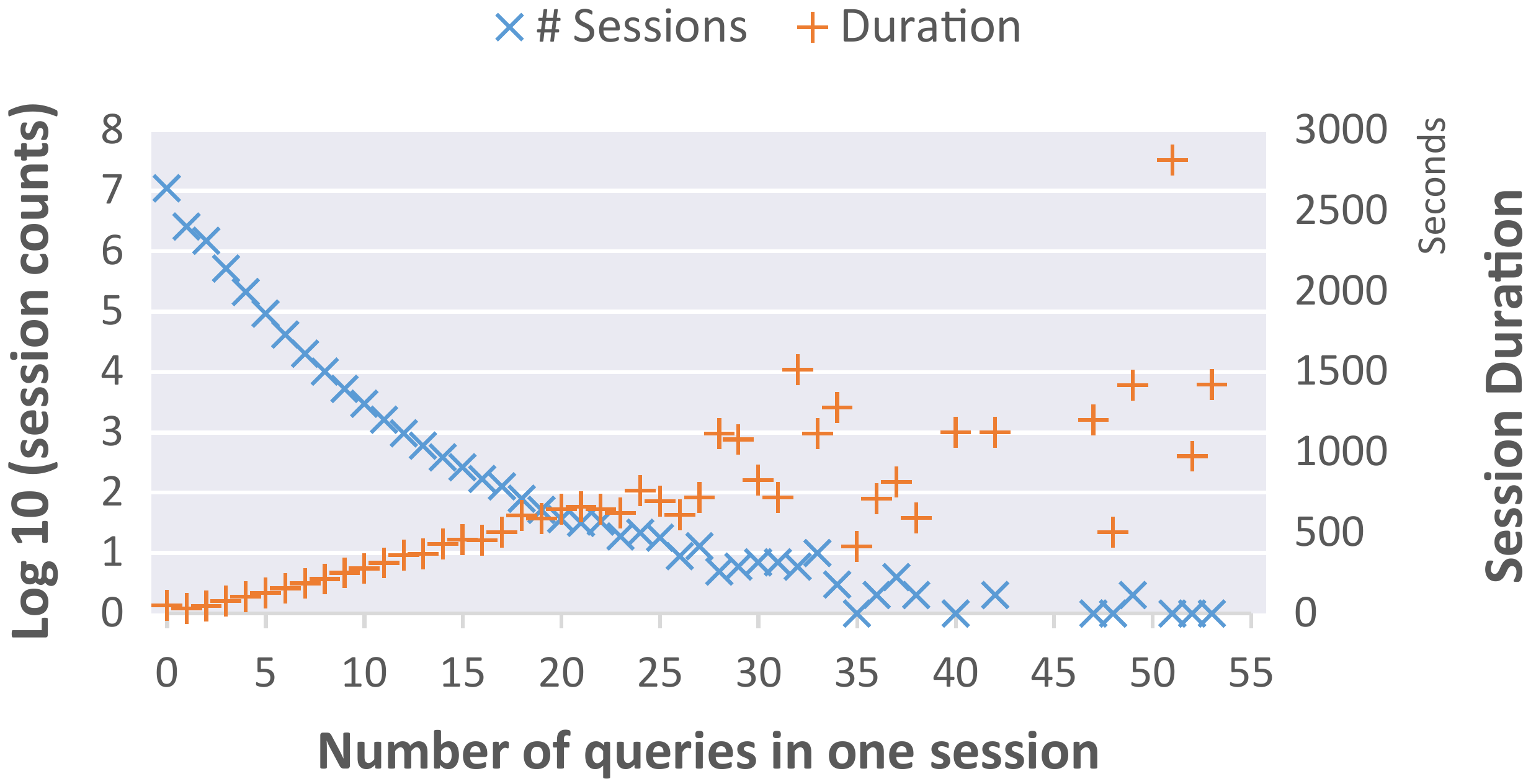} 
    \caption{The distribution of sessions number and duration in terms of the number of queries contained in the session}
    \vspace{-2mm}
    \label{fig:data-querySession}
\end{figure}

From the session perspective, 30.82\% of the sessions contain query event(s), and the users perform an average of 1.82 queries in these sessions.
Fig.~\ref{fig:data-querySession} shows the distribution of the number of sessions and their average session duration in terms of the number of queries contained in the session.
In this figure, we can find as the number of queries in the session increases, the number of such sessions decreases exponentially, while the session duration increases linearly. 

To train our query reformulation model, we obtain the query reformulation records (i.e., the user's process of transforming an original query into a better one) from the user's navigation sequence.
We call these processes the query reformulation threads.
We first remove the sessions that do not contain a query event or of which the last event is not a post \cameraReady{visit}.
The latter limitation is to ensure that the users finish the query reformulation with a desired results. 
\cameraReady{This yields} a dataset containing 8,546,915 events from 3,125,427 sessions generated by 2,216,219 users, in which 4,631,756 events are queries.
We adopt a pattern-based method to extract the query reformulation threads from each session following a greedy approach:

$\cdots, q_1, q_2, \cdots, q_i, p_1(optional), q_{i+1}, \cdots, q_n, p_m \cdots$

In this pattern, we use $q_i$ to denote the $i$-th query event and use $p_i$ to denote the $i$-th post visit event.
All the events are ordered chronologically.
We conjecture that $q_n$ is a relatively better query compared to the queries from $q_1$ to $q_{n-1}$.
Sometimes users may need to visit posts in the query results to determine whether a particular result is what they want, and they may reformulate their queries again after visiting those posts to get better results.
Therefore, to make the pattern more versatile and avoid misrecognition of reformulated queries, post visit event is allowed in the sequence of query events.
More details on extracting query reformulation threads can be found in Section~\ref{sec:approach-collectPair}. 

\section{Empirical Study of Query Reformulation on Stack Overflow}
\label{sec:emp}

In this section, to motivate the required tool support, we perform a formative study to understand the characteristics of query reformulation by analyzing the Stack Overflow log data.

\subsection{What are the characteristics of queries?}
\label{sec:emp-whatQuery}

\noindent\textbf{Query content analysis.}
We analyze the query strings to investigate what the users are searching for on Stack Overflow.
First, we collect all query strings and apply traditional text processing steps (i.e., removing punctuation, transforming to lower case, excluding stop words) to them. 
Then, we identify the most popular $n$-grams in the queries.

Table~\ref{tab:emp-frequentWords} shows the top-10 most frequent 1-grams, 2-grams, 3-grams, and 4-gram in the users' queries.
Programming languages such as \textit{Python} and \textit{Java}, platforms such as \textit{Android}, data types such as \textit{string}, and data structures such as \textit{array} are the most frequently queried terms.
At the same time, \textit{``how to''} is the most frequently used phrase in the queries.
Almost every 3-gram starts with \textit{``how to''}, next comes \textit{``what is''} and programming language qualifiers such as \textit{``in python''} and \textit{``in java''}.
It is worth noting that some \textit{Java} and \textit{Python} error logs appear in the top ten 4-grams.
The reason is that the developers often paste these error logs directly into the search box to perform queries, and logs like ``exception in thread "main"'' and ``ImportError: No module named'' are the most common error types.

\renewcommand\tabcolsep{3.0pt}
\begin{table}[htbp]
\vspace{-2mm}
\footnotesize
\centering
\caption{The top-10 most frequent $n$-grams in the users' queries}
\label{tab:emp-frequentWords}
\begin{tabular}{c|llll}
\toprule
\textbf{Rank} &
\textbf{1-gram} &
\textbf{2-gram} &
\textbf{3-gram} &
\textbf{4-gram} 
\\ 
\midrule

1 & python & how to & how to use & how to create a \\
2 & java & in python & how to get & how to make a \\
3 & file & what is & how do i & how do i use \\
4 & string & in java & how to create & exception in thread "main" \\
5 & android & failed to & how to make & ImportError: No module named \\
6 & c\# & in swift & could not find & how to get the \\
7 & error & in r & how to change & no such file or \\
8 & array & unable to & how can i & how to add a \\
9 & sql & how do & how to find & how to check if \\
10 & list & not found & how to install & such file or directory \\
\bottomrule
\end{tabular}
\vspace{-2mm}
\end{table}

\noindent\textbf{Query length analysis.}
The users may intentionally limit a query's length to avoid returning a few or even empty results when using traditional search engines~\cite{mahdabi2011building}.
To investigate whether this phenomenon exists when the users perform queries on Stack Overflow,
we use whitespaces as the separator to compute the query length (i.e., word count).
Note that we treat words synthesized by CamelCase or underscore\_case as one word since these words often appear in code snippets and can be regarded as the identifiers of variables, classes, or methods.

\begin{figure}[htbp]
    \centering
    \vspace{-1mm}
    \includegraphics[width=0.4\textwidth]{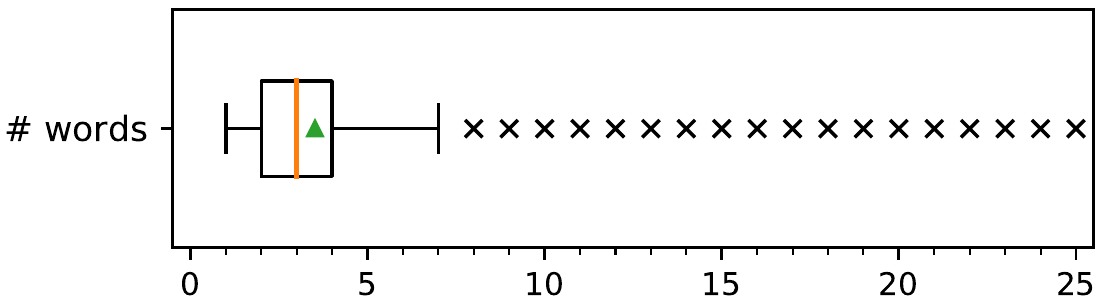} 
    \caption{The distribution of query length by using box plot (note a few outliers with more than 25 words are removed)}
    \vspace{-1mm}
    \label{fig:emp-avgQueryLength}
\end{figure}

Fig.~\ref{fig:emp-avgQueryLength} shows the distribution of query length via a box plot.
The median, mean,  $25^{th}$ percentile, and  $75^{th}$ percentile of the length are 3, 3.6, 2, and 4 respectively.
The distribution shows that users do intentionally limit the length of their queries to get better search results.
However, we can also easily find that the query length span of the outliers is very large, and all values from 8 to 25 correspond to outliers.
For these queries, it is difficult for the search engine of Stack Overflow to return satisfactory results.
Because most search engines are optimized only to handle common requests, they use exact-match techniques in which all query words must match a web page for web page retrieval~\cite{downey2008understanding}. A longer query means lower matching probability and leads to lower-quality search results.
After manual analysis, we find that most of these queries contain error messages or code snippets.

\noindent\textbf{Advanced search methods analysis.}
We notice that some of the queries in the log are structured, which means the users use advanced search methods\footnote{\url{https://stackoverflow.com/help/searching}}.
Advanced search methods can provide users with a convenient way to filter search results.
For example, the users can narrow the search results by only considering the posts with the tag \textit{python} or the posts that have a minimum score of 500.
Stack Overflow provides 29 advanced search methods to help the users filter search results.
We first apply regular expressions to the users' queries to identify the used advanced search method(s).
Then we calculate the ratio of the queries using advanced search and the proportion of each advanced search method.

The result shows that 8.74\% of the queries use the advanced search. 
\cameraReady{The top-10 most frequently used advanced search methods, an illustrative example, and their proportions can be found in Table~\ref{tab:emp-advancedSearch}.}
Tag filtering, user filtering, and declaring specific phrases are the top-3 most commonly-used advanced search methods, accounting for 73.53\%, 10.35\%, and 10.00\% respectively.

\renewcommand{\arraystretch}{1.3}
\begin{table}[htbp]
\footnotesize
\vspace{-1mm}
\centering
\caption{The top-10 most frequently used advanced search methods}
\label{tab:emp-advancedSearch}
\begin{tabular}{llr}
\toprule

\textbf{Type}&
\textbf{Example} &
\textbf{Proportion}
\\
\midrule

tag &
\textbf{{[}powershell{]}} job output &
73.53\% \\

user &
\textbf{user:}8945947 &
10.35\% \\

declare specific phrase &
ios \textbf{"save to files"} &
10.00\% \\

exclude phrase &
accordion \textbf{-jquery} &
3.88\% \\

Wildcard &
\textbf{{[}xamarin*{]}} does not support... &
0.74\% \\

question only &
\textbf{is:question} powershell version &
0.44\% \\

\#answers & 
\textbf{answers:0} firebase &
0.28\% \\

multiple tags &
\textbf{{[}scipy{]} or {[}numpy{]}} array vs matrix &
0.26\% \\

score &
safari cache \textbf{score:3} &
0.19\% \\

creation date&
oauth read gmail \textbf{created:05-04-2015..} &
0.06\% \\

\bottomrule
\end{tabular}
\vspace{-5mm}
\end{table}
\renewcommand{\arraystretch}{1.0}

\subsection{Why are queries reformulated?}
\label{sec:emp-whyreformulate}
The users reformulate their queries for many reasons.
For example, the queries may not specify the corresponding programming language, or some words in the queries may be misspelled.
We use the provided dataset to analyze why the users reformulate their queries.
We manually classify 384 randomly collected query reformulation threads\footnote{\cameraReady{The number is the minimum number to be statistically representative of a large dataset with a confidence level of 95\% and error margin of 5\% via a commonly-used sampling method~\cite{singh2013elements}.}} into four categories.
The first and second author classified these 384 threads independently.
For the cases without an agreement, the final classification result is determined through discussions.
For the threads with more than two queries, if multiple query reformulation types are identified, the type of reformulation from the first query to the last query is used as the final classification result.
The Kappa inter-rater agreement~\cite{viera2005understanding} is 0.83, which shows the high agreement of classification.

Table~\ref{tab:emp-whyReformulate} shows the manual classification results. 
In this table, we can see that adding new information to the query is the most common query reformulation operation, which accounts for 40.1\% of all threads, then comes modifying the query (33.59\%), and deleting information from the query (23.18\%).
In the category of adding new information~\cite{chen2018data}, we further divide it into two sub-categories:
(1) adding information about specific programming languages or platforms,
(2) adding new requirements for the question or more detailed content.
In the category of modifying, we further divide it into three sub-categories:
(1) spelling and syntax checking~\cite{chen2017community},
(2) simplifying and refining the query to reformulate it into the most commonly-used expression,
(3) modifying to other content related to the original query.
In the category of deleting, we further divide it into three sub-categories:
(1) deleting some unnecessary or less informative words,
(2) deleting specific information in error messages or code snippets (such as file path, URL, function names, etc.),
(3) deleting punctuation and some mistyped symbols.

\renewcommand{\arraystretch}{1.2}
\begin{table*}[htbp]
\footnotesize
\vspace{-1mm}
\centering
\caption{Categories of query reformulation}
\label{tab:emp-whyReformulate}
\begin{tabular}{cclc}

\toprule
\textbf{Category} &
\textbf{Sub-category} &
\textbf{Example} &
\textbf{Proportion} \\
\midrule

\multirow{2}{*}{Add} &

Software or platform &
why to use sha1 $=>$ why to use sha1 \textbf{in android}
&
21.61\% \\ & 

Detailed requirement &
db file $=>$ \textbf{open} db file
&
18.49\% \\ 

\cline{1-4}
\multicolumn{3}{r}{\textbf{Category subtotal}} &
40.10\% \\ 
\cline{1-4} 

\multirow{3}{*}{Modify} &

Spelling and syntax check &
.net string.\textbf{emptyp} $=>$ .net string.\textbf{empty}
& 
19.01\% \\ &

Simplify and refine &
\textbf{list inside list} for sightly $=>$ \textbf{nested list} in sightly 
&
12.50\% \\ & 

Turn to related information &
python program \textbf{freezes} $=>$ python program \textbf{hangs}
& 
2.08\% \\ 

\cline{1-4}
\multicolumn{3}{r}{\textbf{Category subtotal}} &
33.59\% \\ 
\cline{1-4} 

\multirow{3}{*}{Delete} & 

Detailed or unnecessary words & 
C\# update a \textbf{keypairvalue} in a Dictionary $=>$ C\# update Dictionary
&
15.10\% \\ & 

Specific information in error message &
\tabincell{l}{\textbf{Property 'getData'} does not exist on type 'ReactInstance' $=>$ \\ does not exist on type 'ReactInstance'}
&
6.51\% \\ & 

Symbols or web links & 
\textbf{:=} dbms\_datapump.open $=>$ dbms\_datapump.open
&
1.56\% \\ 

\cline{1-4} 
\multicolumn{3}{r}{\textbf{Category subtotal}} & 
23.18\% \\ 
\cline{1-4}

Others &  &  & 3.13\% \\ 
\bottomrule
\end{tabular}
\vspace{-3mm}
\end{table*}
\renewcommand{\arraystretch}{1.0}

Based on the classification result, we can observe many different types of query reformulation operations (such as adding information, deleting information, and modifying their expressions).
Previous studies~\cite{ooi2015survey,croft1987approaches,wang2008mining,sisman2013assisting} mainly use rule-based methods to perform automated query reformulation.
However, each of these methods can only target query reformulation for a specific reason. Different rules need to be designed for each situation and then implemented with different methods, which is inefficient and difficult to achieve.
For example, for the query reformulation in the category of modifying, the user may completely change the expression of a query (e.g., from ``\textit{how to cut youtube embedded videos}'' to ``\textit{how to make embedded videos that only play certain parts}'').
It is challenging to implement this type of reformulation by using rule-based methods.
Therefore, we find it necessary to propose a general query reformulation approach, which is the motivation of this study.

\subsection{What is the scale of changes that query reformulation involves?}
\label{sec:emp-scaleChange}
In Section~\ref{sec:emp-whyreformulate}, we find that when reformulating a query, the users may add, remove, or replace some words in the original query.
To understand the scale of changes that query reformulation involves, we measure the similarity between the original query and the reformulated query.
For each query reformulation thread, $n-1$ pairs of reformulation samples ($\textit{original}$, $\textit{reformulated}$) can be extracted, that is $\{(q_1,q_n)$, $(q_2,q_n)$, $\cdots$, $(q_{n-1},q_n)\}$.
The first query $\textit{original}$ in the pair is the initial query performed by the user, and the second query $\textit{reformulated}$ is the user's manual reformulated query, which meets their requirement.
We use a text-matching algorithm with improved dynamic programming~\cite{bergroth2000survey} to find the character-level Longest Common Subsequence (LCS) between $\textit{original}$ and $\textit{reformulated}$. 
Then the similarity between $\textit{original}$ and $\textit{reformulated}$ can be defined as follows:

\begin{equation}
similarity(original, reformulated) = \frac{2\times N_{match}}{N_{total}}
\label{equ:emp-reformulationSimilarity}
\end{equation}
where $N_{match}$ is the number of characters in the LCS, and $N_{total}$ is the sum of the number of characters in $\textit{original}$ and $\textit{reformulated}$.
The similarity score is in the range of 0 to 1.
The higher the similarity score, the fewer changes between $\textit{original}$ and $\textit{reformulated}$.

\begin{figure}[htbp]
    \centering
    \vspace{-2mm}
    \includegraphics[width=0.4\textwidth]{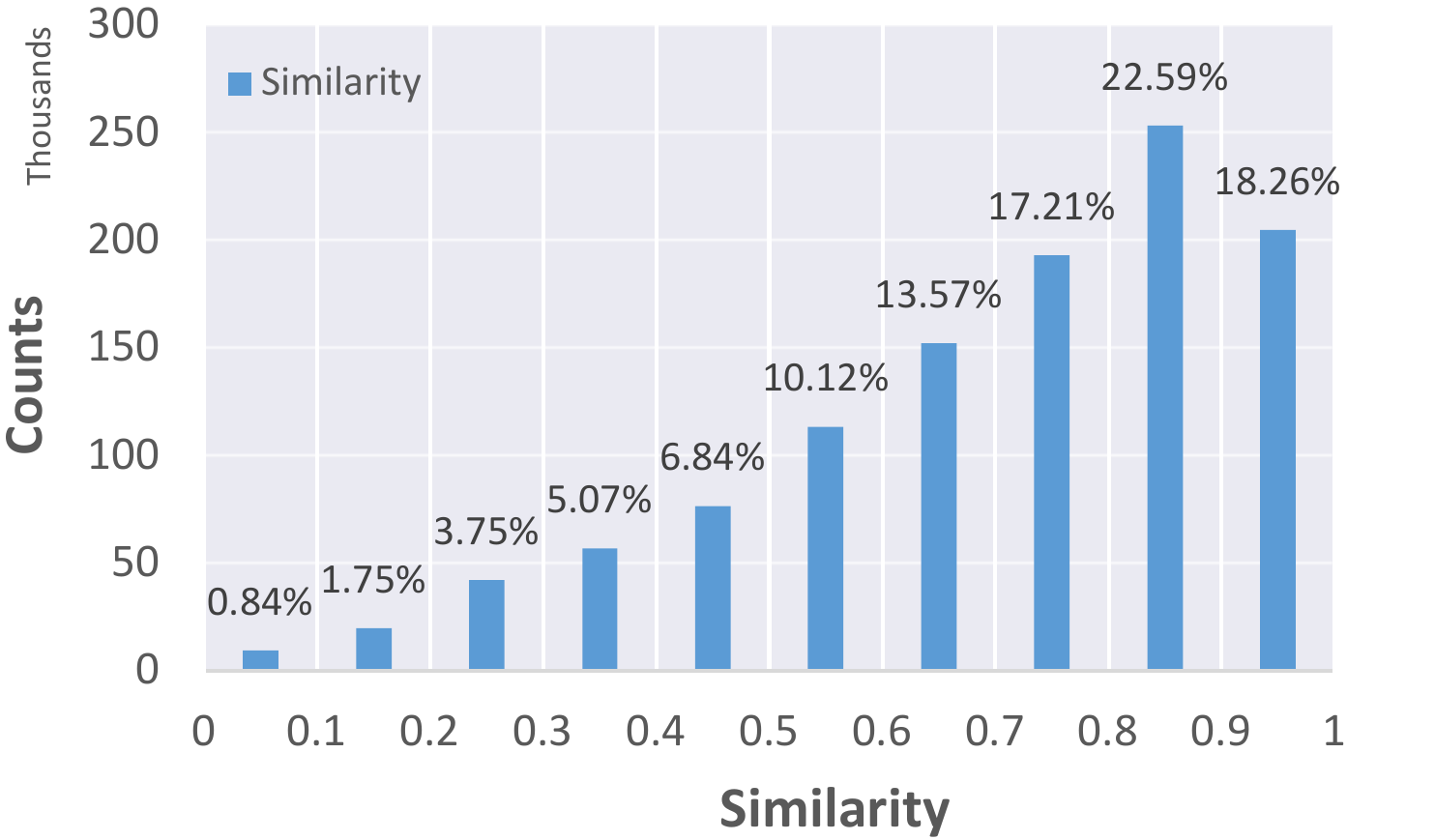} 
    \caption{Distribution of the similarity score of query reformulation pairs}
    \vspace{-6mm}
    \label{fig:emp-reformulationScale}
\end{figure}

As shown in Fig.~\ref{fig:emp-reformulationScale}, among 1,121,185 query reformulation pairs, 58.07\% of the pairs are very similar (i.e., the similarity score is larger than 0.7) with an average of 6.73 character level modifications, which corresponds to 1.14 words according to the average length of query words.
The analysis result indicates that most of the query reformulations only involve minor changes.

\subsection{Summary and implications}
\label{sec:emp-summary}
Our empirical study shows that: 
(1) On Stack Overflow, most of the users' queries are simple and short with only 2 to 4 words, but there are also very long queries containing error logs, code snippets, and structured queries with advanced search patterns.
(2) The users perform query reformulation for various reasons, making it difficult to design a general query reformulation model by only using rule-based methods.
(3) Most query reformulations only involve minor changes.

Considering the diversity of query reformulation patterns, it would require significant effort to manually develop a complete set of reformulation patterns, which is \cameraReady{both} time-consuming and error-prone.
Unlike enumerating all the rules, the deep learning method~\cite{young2018recent} can automatically learn the latent features from the \cameraReady{dataset and these latent features are helpful for downstream tasks.}

Based on the above findings, we believe it is necessary and feasible to propose an automated query reformulation approach based on deep learning.
Our proposed approach would benefit the developers and the Stack Overflow community.
In particular, the developers can use our automated query reformulation tool to refine their queries and get better search results, which can alleviate the effort of manual reformulation.
The Stack Overflow community can also use the query reformulation model to enhance the users' search experience.

\section{Automated Query Reformulation}
\label{sec:approach}

\subsection{Overview of {\tool}}
\label{sec:approach-overview}
Our formative study has shown a wide variety of query reformulation behaviors.
Since not all reformulation patterns are amenable to automated tool support, we focus on query reformulations that trigger minor changes to the original query to develop our automated query reformulation approach. 
In this work, we formulate query reformulation as a machine translation problem, in which an original query is ``translated'' into a reformulated query. 
We solve the problem using the Transformer model~\cite{vaswani2017attention}.

\begin{figure*}[htbp]
    \centering
    \vspace{-2mm}
    \includegraphics[width=0.95\textwidth]{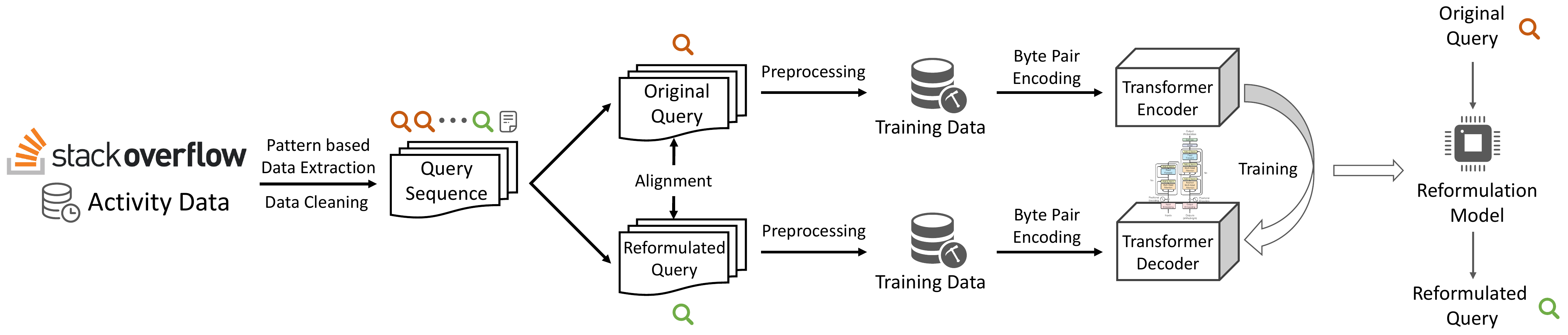} 
    \caption{The overall workflow of our approach {\tool}}
    \vspace{-3mm}
    \label{fig:approach-overview}
\end{figure*}

The overall workflow of {\tool} is shown in Fig.~\ref{fig:approach-overview}.
{\tool} first extracts query reformulation threads for sampling original queries and corresponding reformulated ones.
Then a large corpus of query reformulation pairs is gathered for training the model, each of which contains the original query and the corresponding reformulated one that triggers only minor changes (in Section~\ref{sec:approach-collectPair}).
To model the patterns of query reformulation (such as spelling correction, expression refinement, unnecessary word deletion), {\tool} trains a Transformer-based model with a large parallel corpus of query reformulation pairs (from Section~\ref{sec:approach-bytePairEncoding} to Section~\ref{sec:approach-beamSearch}).
Given the users' original queries, \cameraReady{the trained model} can suggest a list of reformulated candidates for selection.

\subsection{Collection of query reformulation pairs}
\label{sec:approach-collectPair}

Based on the description in Section~\ref{sec:data}, query reformulation threads can be extracted from the users' navigation sequences by pattern matching.
The most important thing for the pattern is how to ensure that $q_i$ and $q_{i+1}$ are issued for the same purpose, rather than two independent queries.
Two constraints are applied:
(1) The character-level similarity between $q_i$ and $q_{i+1}$ must be greater than 0.7 by using Equation~(\ref{equ:emp-reformulationSimilarity}).
(2) The browsing time of the post $p_1$ cannot be greater than 30 seconds.
The first constraint can limit the change scale since a large-scale change is likely to introduce a new query.
The second constraint can guarantee the post $p_1$ is not what the user wants, since a previous study~\cite{liu2010understanding} shows for an unsatisfactory post, the user would not spend more than 30 seconds on it.

We set the minimum number of query events before the final post visiting event in the thread to 2.
After identifying all the query reformulation threads, we perform the following data cleaning steps.
For consecutive identical queries in the thread, we only keep one of these identical queries and remove the rest to avoid generating duplicate query reformulation pairs.
We also remove threads containing queries with non-English words, since non-English words are not encouraged to be used on Stack Overflow\footnote{\url{https://stackoverflow.blog/2009/07/23/non-english-question-policy/}}.
After the above data cleaning, we pair each query between $q_1$ to $q_{n-1}$ with $q_n$ (i.e., ($q_1$, $q_n$), ($q_2$, $q_n$), $\cdots$, ($q_{n-1}$, $q_n$)) as the training instances.
From the provided dataset, we extract a total of 1,121,185 query reformulation pairs.

\cameraReady{Some of these reformulation pairs are hard-to-predict due to their large-scale modifications.
After randomly sampling reformulation pairs with different similarity, we carry out a pilot study to determine the similarity threshold, which can be used to identify large-scale modifications.
The results show that reformulation pairs with a similarity score higher than 0.7 are predictable, while the remaining pairs (i.e., reformulating by large-scale modifications) cannot be predicted even by humans.
As analyzed in Section~\ref{sec:emp-scaleChange}, 58.06\% of the query reformulations involve minor modifications with a similarity of at least 0.7.}
Therefore, we only consider these similar-enough query reformulations, which results in 651,036 query reformulation pairs.

\subsection{Byte pair encoding}
\label{sec:approach-bytePairEncoding}

Traditional word segmentation methods include whitespace-separated word level and character level segmentation.
However, word-level embedding cannot handle the Out-Of-Vocabulary (OOV) problem~\cite{cao2016joint}, and the character level is too fine-grained, resulting in the loss of high-level information.
Therefore, we adopt the BPE (Byte Pair Encoding)~\cite{sennrich2015neural}, which can effectively interpolate between word-level inputs for frequent symbol sequences and character-level inputs for infrequent symbol sequences.

The BPE algorithm contains multiple iterations.
In each iteration, it calculates each consecutive byte pair's frequency and finds the most frequent one, then merges the two byte pair tokens into one token.
This encoding approach can divide words into sub-words like encoding the common words at the word level while encoding the rare words at the character level. 
Since the form of the users' queries on Stack Overflow varies, using BPE can better identify the content of the users' queries.
For example, misspelled words can be divided into several correctly spelled sub-words to alleviate the OOV problem's impact.
\cameraReady{Words with camel-case style can be separated into several sub-words and then each sub-word can be identified.}
Besides, using BPE to separate words with their affixes can help the model learn the relationships between them.
\cameraReady{After applying the BPE algorithm on the corpus, we get a vocabulary composed of sub-words, which is used as the dictionary for our model.}

\subsection{Transformer}
\label{sec:approach-transformer}

With the introduction of the attention mechanism~\cite{bahdanau2014neural}, many neural machine translation approaches integrate the attention mechanism with sequence transduction models like Convolution Neural Network (CNN) and Recurrent Neural Network (RNN) to improve their performance.
\cameraReady{Even so, the CNN-based network architectures require many layers to capture long-term dependencies, leading to high computational cost, and operations in RNN-based network structures cannot be parallelized, resulting in low efficiency.}
To address these issues, Transformer~\cite{vaswani2017attention} is proposed, which \cameraReady{is the first transduction model entirely relying  on the attention mechanism} and has shown competitive performance on various tasks (such as machine translation~\cite{vaswani2017attention}, image captioning~\cite{10.1145/3377811.3380327}, document generation~\cite{liu2018generating}, and syntactic parsing~\cite{kitaev2018constituency}).

The most significant difference between Transformer and \cameraReady{other} sequence transduction models like CNNs and RNNs is that it relies entirely on self-attention (called ``Scaled Dot-Product Attention'') to obtain global dependencies, \cameraReady{and the attention weights are defined by how each word of the sequence is influenced by all the other words in the sequence}.
The self-attention mechanism creates shortcuts between the current token and all the other context tokens to determine the current token vector for the final input representation.
\cameraReady{As weights of these shortcuts are customizable, the self-attention mechanism is able to capture global dependencies without using many layers of convolution and pooling in CNN-based models.
At the same time, the calculation in self-attention is implemented with highly optimized matrix multiplication, and thus resolves the low efficiency caused by the RNN-based models, which sequentially encode the input tokens.}
Intuitively, for an input sequence, first, a neural network is employed to map the input into three matrices: query $Q$, key $K$, and value $V$.
Then, the dot products of the queries $Q$ with all keys $K$ is divided by $\sqrt{d_k}$ ($d_k$ is the dimension of the queries), and a softmax function is applied to obtain the weights for the values.
Finally, the weighted value is used as the representation of each input.

\begin{equation}
\text{ Attention }(Q, K, V)=\operatorname{softmax}\left(\frac{Q K^{T}}{\sqrt{d_{k}}}\right) V
\label{equ:attention}
\end{equation}

Instead of performing a single self-attention function, the Transformer employs a multi-head self-attention, which linearly projects the queries $Q$, keys $K$ and values $V$ $h$ times with different, learned linear projections respectively, where $h$ is the number of heads.
To obtain the temporal relationship of the words, the Transformer adds positional embedding to the input embedding.

\cameraReady{Self-attention can be regarded as a basic calculation in Transformer.}
The Transformer model comprises an encoder and a decoder, which are actually multiple identical encoder and decoder blocks stacked on top of each other with the same number of units.
Each encoder block has one layer of multi-head self-attention followed by another layer of Feed Forward Network (FFN).
On the other hand, each decoder has an extra masked multi-head self-attention, which prevents the model from seeing the generated words during parallel training.
On the encoder side, the multi-head self-attention layer's input is the input embedding with temporal information, and the layer output is normalized and sent into an FFN, which consists of two linear transformations with a ReLU activation.
The output of the encoder on the top of the stack is a set of attention vectors $K$ and $V$, which are used by the decoder to determine the token it should pay attention to.
On the decoder side, the previous output is used as the input to the masked multi-head self-attention layer.
After that, another multi-head self-attention layer with subsequent FFN generates decoder output $h_t$ by getting the query matrix $Q$ from the masked multi-head self-attention layer, the key $K$ and value $V$ matrices from the output of the encoder stack.
Finally, the output of the decoder stack $h_t$ is sent to a fully connected neural network to get the logits vectors, and then a softmax layer to predict the probabilities of the next token.

\begin{equation}
P(w_{t+1}|w_1, \cdots, w_t)=\operatorname{softmax}(h_t W + b)
\label{equ:transformer-output}
\end{equation}
where $h_t$ is the output of the decoder stack.

\subsection{Beam search}
\label{sec:approach-beamSearch}

During decoding, for each time step $t$, the Transformer model will output the word with the highest conditional probability $y_{t}=\operatornamewithlimits{argmax}_{y \in \mathcal{D}} P\left(y | y_{1}, \cdots, y_{t-1}\right)$ via greedy search from $|\mathcal{D}|$ number of words, where $\mathcal{D}$ represents all the words in the word dictionary.
Since we calculate the conditional probability of generating an output sequence based on the input sequence $\prod_{t=1}^{T} P\left(y_{t} | y_{1}, \cdots, y_{t-1}\right)$, where $T$ is the maximum length of the output sequence, the main problem with greedy search is that there is no guarantee that the optimal sequence will be obtained.
The reason is that although the greedy strategy ensures that the output candidate with the highest probability is picked up for each time step, it cannot ensure that the conditional probability of the entire output sequence obtained is the highest.

Beam search~\cite{koehn2004pharaoh} is an improved algorithm of greedy search, and beam size $k$ is its hyper-parameter.
The decoding process using beam search is as follows:
At time step 1, $k$ words with the highest probability are selected as the first word of $k$ candidate output sequences $cs_1$.
Then, at time step $i$, $k$ output sequences with the highest conditional probability $cs_i$ will be selected from $k|\mathcal{D}|$ possible output sequences based on $cs_{i-1}$.
\cameraReady{After $T$ iterations, $k$ sequences with the highest score will be selected from $CS$ as the beam search result, where $T$ is the maximum number of tokens of the output sequence, and $CS$ is the collection from $cs_1$ to $cs_i$.
Note, for each sequence, portions including and after special end-of-sequence tokens are discarded.
}
The score is calculated as follows:

\begin{equation}
\frac{1}{L^{\alpha}} \log P\left(y_{1}, \cdots, y_{L}\right)=\frac{1}{L^{\alpha}} \sum_{t=1}^{L} \log P\left(y_{t} | y_{1}, \cdots, y_{t-1}\right)
\end{equation}
where $L$ is the length of the sequence in $CS$ and $\alpha$ is the length normalization parameter.

\subsection{Implementation}
\label{sec:approach-implementation}

In our implementation, the maximum vocabulary size for BPE~\cite{sennrich2015neural} is set to 10,000.
For the Transformer, we use the tensor2tensor library~\cite{tensor2tensor} developed by the Google Brain team.
The Transformer model contains four attention heads with four encoder and decoder layers, with $\textit{hidden\_size} = 512$.
During the model training, the parameters are learned by back propagation~\cite{werbos1990backpropagation} with Adam optimizer~\cite{kingma2014adam} to minimize the error rate.
We train our model with $\textit{batch\_size} = 256$, $\textit{learning\_rate} = 0.0001$ for 147 epochs on 4 Nvidia V100 GPU (32G memory) for about 8 hours.
During decoding, the hyper-parameter $k$ of beam search is set to 10 to ensure the probability of finding the optimal solution.
The length normalization parameter $\alpha$ is set to 0.6, which is a common practice in neural machine translation~\cite{wang2019learning}.

\begin{figure}
    \centering
    \vspace{-2mm}
    \includegraphics[width=0.45\textwidth]{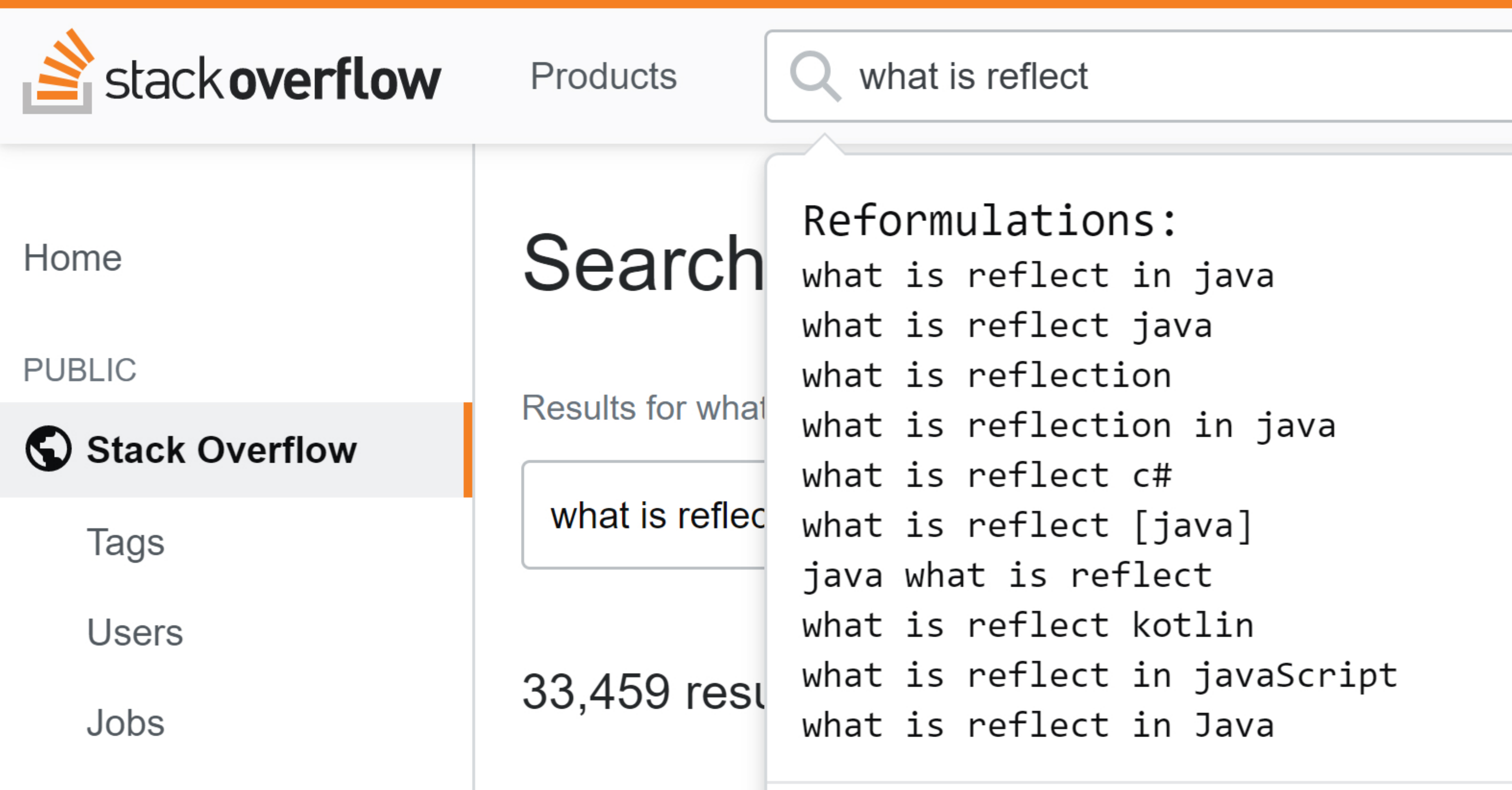} 
    \caption{A screenshot of our query reformulation plugin for the search engine of Stack Overflow}
    \vspace{-5mm}
    \label{fig:approach-plugin}
\end{figure}

To make our work more practical, we develop a browser plugin\footnote{\github} based on Tampermonkey, which is a popular user-script manager.
The browser plugin will automatically analyze the query content and recommend the top-10 query reformulation candidates to the users for selection (a screenshot can be found in Fig.~\ref{fig:approach-plugin}).
Although the plugin is designed to only work on Stack Overflow now, it can be easily extended to other software-specific Q\&A sites.

\section{Quality of Recommended Query Reformulation}
\label{sec:eval}

\subsection{Dataset}
\label{sec:eval-dataset}
From the 651,036 query reformulation pairs, we randomly take 520,830 (80\%) of these query pairs as the training set to train the model, 65,103 (10\%) as the validation set to tune model hyper-parameters, and 65,103 (10\%) as the testing set to evaluate the quality of recommended reformulations.

\subsection{Baselines}
\label{sec:eval-baselines}
As analyzed in Section~\ref{sec:emp-whyreformulate}, many queries are reformulated to fix grammatical errors.
Therefore, we first adopt a popular grammatical error correction (GEC) tool as the baseline.
LanguageTool\footnote{\url{https://languagetool.org}} is an open-source proof-reading tool for more than 20 languages.
This tool's style and grammar checker is rule-based and has been developed for over ten years.

As our task is query reformulation, and Google is the most popular search engine in the world, we choose Google Prediction Service (GooglePS)~\cite{cornea2014providing} as a baseline.
Google uses a prediction service to help complete searches in the search box or address bar within Chrome.
These suggestions are based on the real searches that happen on Google.
Common and trending queries relevant to the strings entered by the users are shown in the drop-down bar for the users to choose\footnote{https://www.blog.google/products/search/how-google-autocomplete-works-search}.
Since Google has accumulated a large-scale dataset of the users' queries, GooglePS can efficiently and accurately reformulate users' queries, such as correcting misspelled words, completing words that the users are typing, \cameraReady{and appending the next possible word}.

The query reformulation task can also be regarded as a translation task (i.e., translating the original query into the reformulated query).
Therefore, we take the most classical neural machine translation model seq2seq~\cite{sutskever2014sequence} as our baseline.
It contains one Long Short-Term Memory (LSTM) model~\cite{hochreiter1997long} as the encoder, which can encode the original query to an embedding vector, and another LSTM model, which can decode that embedding vector to the reformulated query.
We also add the seq2seq model with the attention mechanism as another baseline.

Besides, there are many studies in information retrieval~\cite{ahmad2019context, jiang2018rin, chen2018attention, ahmad2018multi} about query suggestion.
We select HRED-qs~\cite{sordoni2015hierarchical} as one of our baselines, which is a representative hierarchical and session-based query suggestion model with full source code release. 
HRED-qs is trained with our dataset. In particular, given a query in the session, HRED-qs first encodes the information seen up to the position by a query-level RNN encoder and a session-level RNN encoder.  
Then it uses the following decoder to predict the next query.
To keep the setting of all baselines consistent, we only feed the last query before the reformulated one in the session to HRED-qs.

To make a fair comparison, we employ the same data preprocessing method and Byte Pair Encoding for seq2seq and HRED-qs as {\tool}, and we perform hyper-parameter optimization with grid search~\cite{bergstra2011algorithms}.

\subsection{Evaluation metrics}
\label{sec:eval-metrics}
Query reformulation is similar to the grammar error correction task (i.e., revising some words in the original sentence for generating the target sentence).
Therefore, we evaluate the query reformulation with the metrics used in the GEC task (i.e., $\mathit{GLEU}$, $\mathit{M^2}$ and $\mathit{ExactMatch}$). 

$\mathit{GLEU}$ (General Language Evaluation Understanding)~\cite{napoles2015ground, napoles2016gleu} is a customized metric from $\mathit{BLEU}$ (BiLingual Evaluation Understudy)~\cite{papineni2002bleu}, which is a widely used metric to measure the performance of machine translation approaches.
Since only part of the source sentence will be changed in the GEC task, which is different from the machine translation task, this motivates a small change to $\mathit{BLEU}$ that computes $n$-gram precision over the reference but assigns more weight to $n$-grams that have been correctly changed from the source.
Therefore, compared with $\mathit{BLEU}$, $\mathit{GLEU}$ is more suitable for evaluating query reformulation in our study.

$\mathit{MaxMatch}$ (or $\mathit{M^2}$)~\cite{dahlmeier2012better} is another widely used GEC evaluation metric that computes the sequence of phrase-level edits between a source sentence and a system hypothesis that achieves the highest overlap with the gold-standard annotation.
These edits are scored by $\mathit{precision}$,  $\mathit{recall}$, and $\mathit{F1}$.
Specifically, in the scenario of query reformulation, 
$\mathit{M^2@P}$ represents the proportion of edits of the original query given by an approach that appears in the user's manual reformulation.
$\mathit{M^2@R}$ represents the proportion of edits of the original query by users that are correctly predicted by an approach.
$\mathit{M^2@F1}$ is the harmonic mean of $\mathit{M^2@P}$ and $\mathit{M^2@R}$.

$\mathit{ExactMatch}$ ($\mathit{EM}$) evaluates the probability of a perfect match between the query provided by a specific approach and the user's manually reformulated one.
Since {\tool} uses beam search during decoding, it can suggest multiple reformulations for an original query, for the decoding results of {\tool} with different beam size, we can calculate $\mathit{EM@1}$, $\mathit{EM@5}$, and $\mathit{EM@10}$, where $\mathit{EM@n}$ means one case will be considered positive as long as one of $n$ reformulation results returned by beam search matches the ground truth.

\subsection{Evaluation results}
\label{sec:eval-results}
We report our evaluation results by answering the following two research questions.

\subsubsection{RQ1: Can our approach {\tool} generate better reformulated queries than the baselines?}\

Table~\ref{tab:eva-result} shows the evaluation results of query reformulation on the testing set of {\tool} and baselines in terms of all the evaluation metrics.
{\tool} outperforms all state-of-the-art baselines by significant margins in terms of all metrics.
Compared with the best baseline seq2seq (with attention), {\tool} achieves 5.6\% and 4.75\% improvement in terms of $\mathit{EM@10}$ and $\mathit{GLEU}$.
To better illustrate our results, Table~\ref{tab:eva-example} lists examples of query reformulations by different approaches, and more examples can be found on our project site\footnote{\github}.

\renewcommand\tabcolsep{1.5pt}
\renewcommand{\arraystretch}{1.2}
\begin{table}[htbp]
\footnotesize
\vspace{-1mm}
\centering
\caption{Performance of automated query reformulation}
\label{tab:eva-result}
\begin{tabular}{crrrrrrr}
\toprule
Approach & EM@1 & EM@5 & EM@10 & GLEU & M$^2$@P & M$^2$@R & M$^2$@F1 \\
\midrule
LanguageTool & --- & --- & 5.87 & 53.27 & 15.31 & 6.11 & 8.73 \\
GooglePS & 6.20 & 8.14 & 8.58 & 61.07 & 24.67 & 25.84 & 25.24 \\
HRED-qs & 10.40 & 17.63 & 20.01 & 56.85 & 31.80 & 25.82 & 28.50 \\
seq2seq & 11.02 & 23.11 & 28.23 & 61.30 & 36.30 & 25.62 & 30.04 \\
seq2seq+Attn. & 14.53 & 28.47 & 33.77 & 62.93 & 35.93 & 21.15 & 26.62 \\
{\tool} & \textbf{22.21} & \textbf{33.47} & \textbf{39.37} & \textbf{67.68} & \textbf{39.67} & \textbf{31.97} & \textbf{35.41} \\ 
\bottomrule
\end{tabular}
\vspace{-1mm}
\end{table}
\renewcommand{\arraystretch}{1.0}

\renewcommand{\arraystretch}{1.0}
\renewcommand\tabcolsep{1pt}
\begin{table*}[htbp]
\fontsize{6.5}{6.5}\selectfont
\vspace{-2mm}
\centering
\caption{Examples of query reformulation by different approaches (``---'' represents no reformulation suggestions)}
\label{tab:eva-example}
\begin{tabular}{ll|l|l|l}
\toprule

\textbf{ID} &
\multicolumn{1}{c}{\textbf{Original Query}} &
\multicolumn{1}{c}{\textbf{GooglePS}} &
\multicolumn{1}{c}{\textbf{seq2seq + Attn.}} &
\multicolumn{1}{c}{\textbf{{\tool}}} \\
\midrule

1 &
pandas delete last characters in strin &
\textbf{python} delete last characters in \textbf{string} &
pandas delete last characters in \textbf{string} &
\textbf{python} delete last characters in \textbf{string}
\\\rowcolor{gray!20}

2 &
playing sound in swift3 &
playing sound in \textbf{swift 3} &
\multicolumn{1}{c|}{---} &
\textbf{play} sound in \textbf{swift 3}
\\

3 &
swift grab currentdate &
\multicolumn{1}{c|}{---} &
swift \sout{grab} \textbf{current date} &
swift \textbf{get} \textbf{current date}
\\\rowcolor{gray!20}

4 &
assign string to number C3 &
\multicolumn{1}{c|}{---} &
assign string to number \textbf{C\#} &
assign string to number \textbf{C\#}
\\

5 &
do and while in java &
do and while \textbf{loop} in java &
\multicolumn{1}{c|}{---} &
do and while \textbf{loop} in java
\\\rowcolor{gray!20}

6 &
requests negotiate &
requests negotiate \textbf{python} &
python requests negotiate &
\textbf{[python]} requests negotiate
\\

7 &
/usr/bin/ld: skipping incompatible libpthread.so &
\multicolumn{1}{c|}{---} &
\textbf{lib}skipping incompatible libpthread.so &
\sout{/usr/bin/ld:} skipping incompatible libpthread.so
\\\rowcolor{gray!20}

8 &
subtracting the pandas series rf from all columns &
\multicolumn{1}{c|}{---} & 
subtracting the pandas series \sout{rf from all columns} &
subtracting the pandas series \sout{rf} from all columns
\\

9 &
\textless{}trigger\textgreater{}Missing report definition\textless{}/trigger\textgreater{} &
\multicolumn{1}{c|}{---} &
\sout{\textless{}trigger\textgreater{}}Missing report definition\textless{}/trigger\textgreater{} &
\sout{\textless{}trigger\textgreater{}} Missing report definition \sout{\textless{}trigger\textgreater{}}
\\\rowcolor{gray!20}

10 &
opencv scale image &
opencv scale image \textbf{to 0 1} &
\multicolumn{1}{c|}{---} &
opencv \textbf{resize} image
\\

11 &
a* search &
a* search \textbf{example} &
a* search \textbf{python} & 
a\sout{*} \textbf{star} search
\\\rowcolor{gray!20}

12 &
df -h show disk &
\multicolumn{1}{c|}{---} &
\multicolumn{1}{c|}{---} & 
\textbf{"df -h"} show disk
\\

13 &
resteasy how to support websocket &
\multicolumn{1}{c|}{---} &
resteasy \sout{how to support} websocket &
resteasy \sout{how to support} websocket
\\\rowcolor{gray!20}

14 &
volume control programatically android &
\textbf{android} volume control \textbf{programmatically} &
\multicolumn{1}{c|}{---} &
volume control android
\\

\bottomrule
\end{tabular}
\vspace{-6mm}
\end{table*}
\renewcommand{\arraystretch}{1.0}

GEC tools perform well in correcting misspellings (``\textit{strin}'' to ``\textit{string}'' in Example 1), grammar issues (``\textit{playing}'' to ``\textit{play}'' in Example 2), and sentence format (``\textit{currentdate}'' to ``\textit{current date}'' in Example 3).
However, they only achieve a 5.87\% exact match, as spelling errors only account for 19\% of the reasons for the users' query reformulations (in Section \ref{sec:emp-whyreformulate}).
In addition, they have difficulty in detecting spelling errors specific to the programming field.
For example, they cannot detect any misspelled words in ``{\footnotesize how to import \underline{bumpy} array}'' where the word ``\textit{bumpy}'' is the wrong spelling of ``\textit{numpy}'' (a \textit{Python} library).
Instead, {\tool} can correct these software-specific misspelled words by learning the domain knowledge from our software-specific dataset (another example is from ``\textit{C3}'' to ``\textit{C\#}'' in Example 4).

GooglePS can perform complex reformulations by learning from billions of queries Google processes every day, such as adding important missing keywords (``\textit{loop}'' in Example 5) and suggesting language/platform limitations (``\textit{python}'' in Examples 1 and 6).
However, as a general-purpose search engine, Google does not perform well in software-specific query reformulation, especially for unpopular software-specific queries.
For example, GooglePS cannot reformulate the query ``{\footnotesize /usr/bin/ld: skipping incompatible libpthread.so}'' in Example 7, but {\tool} can remove the file directory to make it more general.
As the file directory often varies between developers' coding environments, it should be removed to keep the query more general, for retrieving more accurate results.
By learning from massive query reformulations, which are software-specific from Stack Overflow, {\tool} can effectively revise such queries by applying software-specific reformulation strategies (similar examples in Examples 8 and 9).

Since using the same training data as {\tool}, the HRED-qs, seq2seq and seq2seq with attention model can capture software-specific semantics during query reformulation.
This is the reason why they achieve better performance than the other baselines.
However, since the goal of HRED-qs is basically ``next query prediction'', the context-aware hierarchical encoding does not enhance the ability of the decoder to generate a reformulated query, but may obscure the information of the original query, causing a deviation of the decoder result.
For example, ``{\footnotesize comboox lost focus}'' should be reformulated to ``{\footnotesize combobox lost focus}'', but HRED-qs reformulates it to ``{\footnotesize \underline{iphone x} lost focus}''.
Besides, the problem for both HRED-qs and seq2seq is that the input is encoded into one single vector representation, which may not be sufficient to store all the information, especially for long queries.
For example, ``{\footnotesize Allow user to paste url with .ph and rpeturn clean url}'' should be reformulated to ``{\footnotesize Allow user to paste url with \underline{.php} and \underline{return} clean url}''.
However, seq2seq model can only reformulate it to ``{\footnotesize Allow user to paste url with php}'' due to the query length.
On the contrary, {\tool} does not have this problem by using an attention-based Transformer model.

In addition to the types of query reformulations mentioned above, {\tool} also outperforms baselines in more complex reformulations such as 
revising with more commonly-used software-specific terms (e.g., ``{\footnotesize swift grab currentdate}'' to ``{\footnotesize swift \underline{get} current date}'' in Example 3, ``{\footnotesize opencv scale image}'' to ``{\footnotesize opencv \underline{resize} image}'' in Example 10),
replace symbols with text (e.g., ``{\footnotesize a* search}'' to ``{\footnotesize a \underline{star} search}'' in Example 11) or enclose symbols that will accidentally trigger advanced search in quotation marks (e.g., ``{\footnotesize df -h show disk}'' to ``{\footnotesize \underline{"df -h"} show disk}'' in Example 12 to prevent showing results for ``\textit{df show disk}'' and not containing ``\textit{h}''), and
simplify or refine the query (e.g., ``{\footnotesize resteasy how to support websocket}'' to ``{\footnotesize resteasy websocket}'' in Example 13 and ``{\footnotesize volume control programatically android}'' to ``{\footnotesize volume control android}'' in Example 14).

\subsubsection{RQ2: What types of query reformulations are challenging for {\tool} to deal with?}\

Although {\tool} can achieve the best performance compared to state-of-the-art baselines, {\tool} also makes mistakes in some query reformulations.
We manually check some \cameraReady{randomly sampled} erroneous reformulations and identify two main reasons why our reformulations do not match the ground truth.

First, some queries are edited to add more information, which is beyond the context of the original query such as ``{\footnotesize python covert variable to integer}'' to ``{\footnotesize python \underline{convert} \underline{hexadecimal} to integer}'' and ``{\footnotesize git commit -am}'' to ``{\footnotesize git commit -am \underline{vs git add}}''.
Although {\tool} can successfully revise the misspelling ``\textit{covert}'' to ``\textit{convert}'', it cannot guess the replacement of ``\textit{variable}'' with ``\textit{hexadecimal}'' or adding ``\textit{vs git add}'' by considering only the local context of the query. 
To support such complicated reformulation, we need to consider the broader context of the search (e.g., previous queries) in the future.

Second, the same meaning may be expressed in different ways.
For example, given the original query ``{\footnotesize remove , from input}'', {\tool} recommends the reformulated query as ``{\footnotesize remove \underline{comma} from input}'', however, the ground truth is ``{\footnotesize remove \underline{","} from input}''.
Similarly, given the original query ``{\footnotesize read mouse cursor}'', {\tool} recommends the reformulated query as ``{\footnotesize \underline{c\#} read mouse cursor}'', however, the ground truth is ``{\footnotesize read mouse cursor \underline{in C\#}}''.
Although the users' reformulation results and our recommendations are not exactly matched, they convey the same meaning.
Some of our recommendations are of higher quality than the users' reformulation and may lead to better search results.
That is also why the performance of {\tool} is highly underestimated.

\subsection{Discussion}
\cameraReady{In Section~\ref{sec:eval-results}, we evaluate the quality of reformulations from different approaches by comparing them with users' manual ones with the metrics used in the GEC task.
However, none of these metrics consider the model effectiveness (i.e., the ability of the reformulated query to retrieve the desired post).
Therefore, we further evaluate the retrieval effectiveness of {\tool} in this section.}

\cameraReady{For each query reformulation thread (mentioned in Section~\ref{sec:data}), we collect pairs of users' original query and finally-visited post, which is assumed to be the target post.
These query-post pairs can be used to demonstrate the retrieval effectiveness of the model.
For each pair, we adopt both our approach and baselines in Section~\ref{sec:eval-baselines} to reformulate the query and check the ranking of the target post in all candidate posts on Stack Overflow.
All the 65,103 queries in the testing set are paired with their corresponding finally-visited post as the evaluation data.
}

\cameraReady{We use $\mathit{MRR}$ (Mean Reciprocal Rank) as the metric for retrieval effectiveness evaluation.
$\mathit{MRR}$ is the average of the reciprocal ranks (i.e., the multiplicative inverse of the target post's rank in the search result) of the search results for all the queries.
For example, given a query, if the target post ranks fifth in the search result, the reciprocal rank is 1/5=0.2.
A higher $\mathit{MRR}$ value indicates a better search result.}

\cameraReady{The comparison results between {\tool} and baselines in terms of $\mathit{MRR}$ can be found in Table~\ref{tab:discussion-effectiveness}.
{\tool} achieves the best performance (i.e., 129.33\% boost in terms of $\mathit{MRR}$ to the original query).
Even compared with the best baseline seq2seq+Attn., {\tool} still achieves a 23.7\% boost.
This result shows that the reformulated query given by {\tool} is not only the closest to manual reformulation, but also has the best retrieval effectiveness among all the baselines.
Therefore, {\tool} can effectively help users obtain better search results via high-quality query reformulation.}

\renewcommand{\arraystretch}{1.2}
\renewcommand\tabcolsep{3pt}
\begin{table}[htbp]
\footnotesize
\vspace{-1mm}
\centering
\caption{Evaluation result of retrieval effectiveness}
\label{tab:discussion-effectiveness}
\begin{tabular}{ccc}
\toprule
Approach & $\mathit{MRR}$ & Boost Rate  \\ 
\midrule
Original Query    & 0.075  & ---                     \\
LanguageTool      & 0.071  & 5.33\% $\downarrow$       \\
GooglePS          & 0.083  & 10.67\% $\uparrow$        \\
HRED-qs           & 0.108  & 44.00\% $\uparrow$        \\
seq2seq           & 0.127  & 69.33\% $\uparrow$        \\
seq2seq+Attn.     & 0.139  & 85.33\% $\uparrow$        \\
{\tool}   & \textbf{0.172} & \textbf{129.33\%} $\uparrow$       \\
\bottomrule
\end{tabular}
\vspace{-1mm}
\end{table}
\renewcommand{\arraystretch}{1.0}

\section{Related Work}
\label{sec:relatedwork}
Information Retrieval (IR) has been widely used in software engineering (SE) tasks, such as traceability recovery~\cite{oliveto2010equivalence,mcmillan2009combining}, feature location~\cite{gay2009use,dit2013integrating}, library migration~\cite{chen2016mining,chen2019s,chen2016similartech}, API search~\cite{chen2019mining,chen2020similarapi} and GUI design seeking~\cite{chen2020wireframe,chen2019gallery,chen2020lost}. 
In this section, we summarize the related works about query reformulation in general IR and its application in SE domain.

\subsection{Query reformulation in general information retrieval}
\label{sec:relatedwork-general}
To help users better refine their queries, there are many studies on query expansion~\cite{nie2016query,lu2015query}, query reformulation~\cite{rahman2019automatic,rahman2018effective} and query suggestion~\cite{ahmad2019context, jiang2018rin, chen2018attention, ahmad2018multi} in information retrieval.
In detail, Jiang et al.~\cite{jiang2018rin} tried to provide query suggestions based on previous queries in the session by introducing a binary classifier and an RNN-based decoder as the query discriminator and the query generator.
Chen et al.~\cite{chen2018attention} proposed an attention-based hierarchical neural query suggestion model that combines a session-level neural network and a user-level neural network to model the users' short and long term search history.

However, most of these studies are session-based or user profile-based and apply to general text search.
Different from general text search, search in the software engineering domain is very specific, with domain-specific terms and code snippets, which make general approaches not applicable in this scenario.
Therefore, we carry out this study based on domain-specific dataset for providing a software-specific automated query reformulator.

\subsection{Query reformulation for document search in SE}
\label{sec:relatedwork-documentSearch}
The performance of document search in software engineering relies on the domain-specific query reformulation.
Haiduc et al. proposed several metrics to measure query difficulty~\cite{haiduc2011effect}, query specificity~\cite{haiduc2012evaluating}, and query quality~\cite{haiduc2012automatic} for concept location.
Based on these metrics, they~\cite{haiduc2013automatic,haiduc2013query} further developed a machine learning model to adopt one of four strategies to recommend revised queries. 
Rahman et al.~\cite{rahman2016quickar} proposed a word-embedding based method to extract semantically similar terms from questions on Stack Overflow, hence suggesting semantically relevant queries.
Li et al.~\cite{li2016query} shared a similar idea by building a software-specific domain lexical database based on tags on Stack Overflow and optimized the input queries to help search software-related documents. 
Chen et al.~\cite{chen2016learning} reformulated the Chinese queries to English ones for searching related posts on Stack Overflow.

Most of these previous studies generate query reformulation based on heuristic rules or lexical databases, which depends greatly on the quality and size of the rules or the database.
In contrast, {\tool} is fully data-driven, which is based on large-scale real-world developers' queries on Stack Overflow. 
We believe that {\tool} can automatically learn reformulation patterns, and generate better reformulation result.

\subsection{Query reformulation for code search in SE}
\label{sec:relatedwork-codeSearch}
Code search plays an important role in software engineering, many previous studies focused on query reformulation for code search.
Sisman et al.~\cite{sisman2013assisting} proposed a query reformulation framework by enriching the users' queries with certain specific terms drawn from the highest-ranked retrieved artifacts.
Howard et al.~\cite{howard2013automatically} leveraged similar word pairs in comments and method signatures, which are semantically similar in software engineering to reformulate users' queries.
Lu et al.~\cite{lu2015query} took a similar method, i.e., identifying each term in the original query and extends with synonyms generated from WordNet.
Nie et al.~\cite{nie2016query} identified software-specific expansion words from high-quality pseudo relevance feedback question and answer pairs on Stack Overflow.
Rahman et al.~\cite{rahman2017improved} identified terms from the source code using a novel term weight-CodeRank, and then suggested effective reformulation of the original query by exploiting the source document structures, query quality analysis, and machine learning.
They further proposed a query reformulation technique that suggests a list of relevant API classes for a natural language query by exploiting keyword-API associations from the questions and answers on Stack Overflow~\cite{rahman2018effective}.
Similarly, Sirres et al.~\cite{sirres2018augmenting} augmented the original query with structural code entities by mining questions and answers from Stack Overflow.

Different from code search, our study mainly focuses on software-specific document search.
Moreover, {\tool} can complement these code search approaches to reformulate the users' queries better.

\section{Conclusion}
\label{sec:conclusion}

Constructing an efficient query to search through a large amount of programming knowledge is a challenging task for developers, especially for novices.
Our empirical study on a large scale real-world query records on Stack Overflow indicates that developers always reformulate their queries to obtain the desired results.
To assist with developers' efficient search, we propose a deep learning-based approach {\tool} to learn query reformulation patterns from query logs provided by Stack Overflow.
Given the original query, it can automatically recommend a list of reformulation candidates for selection.
Evaluation on large-scale archival query reformulations verifies the superiority of {\tool} compared with five state-of-the-art baselines. 

In the future, we will further improve the performance of {\tool} by incorporating more contextual information such as the users' profile, query history, and their post visiting history.
Moreover, we plan to take our approach one step further by directly recommending posts for the query.
Based on users' queries and corresponding clicked posts provided by Stack Overflow, we could develop a domain-specific model to learn the relationships between them.

\section*{Acknowledgement}

The authors would like to thank Stack Exchange Inc. for sharing the dataset, and the anonymous reviewers for their insightful comments and suggestions.
This work is supported in part by the National Natural Science Foundation of China (Grant Nos. 61872263, 61702041 and 61202006), the Open Project of State Key Laboratory for Novel Software Technology at Nanjing University (Grant No. KFKT2019B14), and the Australian Research Council (DE180100153). 

\normalem
\bibliographystyle{IEEEtran}
\bibliography{ref}

\begin{thebibliography}{10}
\providecommand{\url}[1]{#1}
\csname url@samestyle\endcsname
\providecommand{\newblock}{\relax}
\providecommand{\bibinfo}[2]{#2}
\providecommand{\BIBentrySTDinterwordspacing}{\spaceskip=0pt\relax}
\providecommand{\BIBentryALTinterwordstretchfactor}{4}
\providecommand{\BIBentryALTinterwordspacing}{\spaceskip=\fontdimen2\font plus
\BIBentryALTinterwordstretchfactor\fontdimen3\font minus
  \fontdimen4\font\relax}
\providecommand{\BIBforeignlanguage}[2]{{%
\expandafter\ifx\csname l@#1\endcsname\relax
\typeout{** WARNING: IEEEtran.bst: No hyphenation pattern has been}%
\typeout{** loaded for the language `#1'. Using the pattern for}%
\typeout{** the default language instead.}%
\else
\language=\csname l@#1\endcsname
\fi
#2}}
\providecommand{\BIBdecl}{\relax}
\BIBdecl

\bibitem{huang2018tell}
Y.~Huang, C.~Chen, Z.~Xing, T.~Lin, and Y.~Liu, ``Tell them apart: distilling
  technology differences from crowd-scale comparison discussions,'' in
  \emph{2018 33rd IEEE/ACM International Conference on Automated Software
  Engineering (ASE)}.\hskip 1em plus 0.5em minus 0.4em\relax IEEE, 2018, pp.
  214--224.

\bibitem{chen2016mining}
C.~Chen and Z.~Xing, ``Mining technology landscape from stack overflow,'' in
  \emph{Proceedings of the 10th ACM/IEEE International Symposium on Empirical
  Software Engineering and Measurement}, 2016, pp. 1--10.

\bibitem{chen2016techland}
C.~Chen, Z.~Xing, and L.~Han, ``Techland: Assisting technology landscape
  inquiries with insights from stack overflow,'' in \emph{2016 IEEE
  International Conference on Software Maintenance and Evolution
  (ICSME)}.\hskip 1em plus 0.5em minus 0.4em\relax IEEE, 2016, pp. 356--366.

\bibitem{abdalkareem2017developers}
R.~Abdalkareem, E.~Shihab, and J.~Rilling, ``What do developers use the crowd
  for? a study using stack overflow,'' \emph{IEEE Software}, vol.~34, no.~2,
  pp. 53--60, 2017.

\bibitem{xia2017developers}
X.~Xia, L.~Bao, D.~Lo, P.~S. Kochhar, A.~E. Hassan, and Z.~Xing, ``What do
  developers search for on the web?'' \emph{Empirical Software Engineering},
  vol.~22, no.~6, pp. 3149--3185, 2017.

\bibitem{chen2016towards}
C.~Chen and Z.~Xing, ``Towards correlating search on google and asking on stack
  overflow,'' in \emph{Proceedings of 2016 IEEE 40th Annual Computer Software
  and Applications Conference}.\hskip 1em plus 0.5em minus 0.4em\relax IEEE,
  2016, pp. 83--92.

\bibitem{datta2008image}
R.~Datta, D.~Joshi, J.~Li, and J.~Z. Wang, ``Image retrieval: Ideas,
  influences, and trends of the new age,'' \emph{ACM Computing Surveys},
  vol.~40, no.~2, pp. 1--60, 2008.

\bibitem{zha2010visual}
Z.-J. Zha, L.~Yang, T.~Mei, M.~Wang, Z.~Wang, T.-S. Chua, and X.-S. Hua,
  ``Visual query suggestion: Towards capturing user intent in internet image
  search,'' \emph{ACM Transactions on Multimedia Computing, Communications, and
  Applications}, vol.~6, no.~3, pp. 1--19, 2010.

\bibitem{chen2017unsupervised}
C.~Chen, Z.~Xing, and X.~Wang, ``Unsupervised software-specific morphological
  forms inference from informal discussions,'' in \emph{2017 IEEE/ACM 39th
  International Conference on Software Engineering (ICSE)}.\hskip 1em plus
  0.5em minus 0.4em\relax IEEE, 2017, pp. 450--461.

\bibitem{chen2019sethesaurus}
X.~Chen, C.~Chen, D.~Zhang, and Z.~Xing, ``Sethesaurus: Wordnet in software
  engineering,'' \emph{IEEE Transactions on Software Engineering}, 2019.

\bibitem{jansen2009patterns}
B.~J. Jansen, D.~L. Booth, and A.~Spink, ``Patterns of query reformulation
  during web searching,'' \emph{Journal of the american society for information
  science and technology}, vol.~60, no.~7, pp. 1358--1371, 2009.

\bibitem{sloan2015term}
M.~Sloan, H.~Yang, and J.~Wang, ``A term-based methodology for query
  reformulation understanding,'' \emph{Information Retrieval Journal}, vol.~18,
  no.~2, pp. 145--165, 2015.

\bibitem{bing2015web}
L.~Bing, W.~Lam, T.-L. Wong, and S.~Jameel, ``Web query reformulation via joint
  modeling of latent topic dependency and term context,'' \emph{ACM
  Transactions on Information Systems (TOIS)}, vol.~33, no.~2, pp. 1--38, 2015.

\bibitem{jiang2014learning}
J.-Y. Jiang, Y.-Y. Ke, P.-Y. Chien, and P.-J. Cheng, ``Learning user
  reformulation behavior for query auto-completion,'' in \emph{Proceedings of
  the 37th international ACM SIGIR conference on Research \& development in
  information retrieval}, 2014, pp. 445--454.

\bibitem{rieh2006analysis}
S.~Y. Rieh \emph{et~al.}, ``Analysis of multiple query reformulations on the
  web: The interactive information retrieval context,'' \emph{Information
  Processing \& Management}, vol.~42, no.~3, pp. 751--768, 2006.

\bibitem{huang2009analyzing}
J.~Huang and E.~N. Efthimiadis, ``Analyzing and evaluating query reformulation
  strategies in web search logs,'' in \emph{Proceedings of the 18th ACM
  conference on Information and knowledge management}, 2009, pp. 77--86.

\bibitem{sutskever2014sequence}
I.~Sutskever, O.~Vinyals, and Q.~V. Le, ``Sequence to sequence learning with
  neural networks,'' in \emph{Proceedings of Advances in neural information
  processing systems}, 2014, pp. 3104--3112.

\bibitem{sordoni2015hierarchical}
A.~Sordoni, Y.~Bengio, H.~Vahabi, C.~Lioma, J.~Grue~Simonsen, and J.-Y. Nie,
  ``A hierarchical recurrent encoder-decoder for generative context-aware query
  suggestion,'' in \emph{Proceedings of the 24th ACM International on
  Conference on Information and Knowledge Management}, 2015, pp. 553--562.

\bibitem{cornea2014providing}
R.~C. Cornea and N.~B. Weininger, ``Providing autocomplete suggestions,''
  Feb.~4 2014, uS Patent 8,645,825.

\bibitem{sadowski2015developers}
C.~Sadowski, K.~T. Stolee, and S.~Elbaum, ``How developers search for code: a
  case study,'' in \emph{Proceedings of the 2015 10th Joint Meeting on
  Foundations of Software Engineering}, 2015, pp. 191--201.

\bibitem{mahdabi2011building}
P.~Mahdabi, M.~Keikha, S.~Gerani, M.~Landoni, and F.~Crestani, ``Building
  queries for prior-art search,'' in \emph{Proceedings of Information Retrieval
  Facility Conference}.\hskip 1em plus 0.5em minus 0.4em\relax Springer, 2011,
  pp. 3--15.

\bibitem{downey2008understanding}
D.~Downey, S.~Dumais, D.~Liebling, and E.~Horvitz, ``Understanding the
  relationship between searchers' queries and information goals,'' in
  \emph{Proceedings of the 17th ACM conference on Information and knowledge
  management}, 2008, pp. 449--458.

\bibitem{singh2013elements}
R.~Singh and N.~S. Mangat, \emph{Elements of survey sampling}.\hskip 1em plus
  0.5em minus 0.4em\relax Springer Science \& Business Media, 2013, vol.~15.

\bibitem{viera2005understanding}
A.~J. Viera, J.~M. Garrett \emph{et~al.}, ``Understanding interobserver
  agreement: the kappa statistic,'' \emph{Fam med}, vol.~37, no.~5, pp.
  360--363, 2005.

\bibitem{chen2018data}
C.~Chen, X.~Chen, J.~Sun, Z.~Xing, and G.~Li, ``Data-driven proactive policy
  assurance of post quality in community q\&a sites,'' \emph{Proceedings of the
  ACM on human-computer interaction}, vol.~2, no. CSCW, pp. 1--22, 2018.

\bibitem{chen2017community}
C.~Chen, Z.~Xing, and Y.~Liu, ``By the community \& for the community: a deep
  learning approach to assist collaborative editing in q\&a sites,''
  \emph{Proceedings of the ACM on Human-Computer Interaction}, vol.~1, no.
  CSCW, pp. 1--21, 2017.

\bibitem{ooi2015survey}
J.~Ooi, X.~Ma, H.~Qin, and S.~C. Liew, ``A survey of query expansion, query
  suggestion and query refinement techniques,'' in \emph{Proceedings of 2015
  4th International Conference on Software Engineering and Computer
  Systems}.\hskip 1em plus 0.5em minus 0.4em\relax IEEE, 2015, pp. 112--117.

\bibitem{croft1987approaches}
W.~B. Croft, ``Approaches to intelligent information retrieval.''
  \emph{Information Processing and Management}, vol.~23, no.~4, pp. 249--54,
  1987.

\bibitem{wang2008mining}
X.~Wang and C.~Zhai, ``Mining term association patterns from search logs for
  effective query reformulation,'' in \emph{Proceedings of the 17th ACM
  conference on Information and knowledge management}, 2008, pp. 479--488.

\bibitem{sisman2013assisting}
B.~Sisman and A.~C. Kak, ``Assisting code search with automatic query
  reformulation for bug localization,'' in \emph{Proceedings of 2013 10th
  Working Conference on Mining Software Repositories}.\hskip 1em plus 0.5em
  minus 0.4em\relax IEEE, 2013, pp. 309--318.

\bibitem{bergroth2000survey}
L.~Bergroth, H.~Hakonen, and T.~Raita, ``A survey of longest common subsequence
  algorithms,'' in \emph{Proceedings Seventh International Symposium on String
  Processing and Information Retrieval. SPIRE 2000}.\hskip 1em plus 0.5em minus
  0.4em\relax IEEE, 2000, pp. 39--48.

\bibitem{young2018recent}
T.~Young, D.~Hazarika, S.~Poria, and E.~Cambria, ``Recent trends in deep
  learning based natural language processing,'' \emph{IEEE Computational
  IntelligenCe Magazine}, vol.~13, no.~3, pp. 55--75, 2018.

\bibitem{vaswani2017attention}
A.~Vaswani, N.~Shazeer, N.~Parmar, J.~Uszkoreit, L.~Jones, A.~N. Gomez,
  {\L}.~Kaiser, and I.~Polosukhin, ``Attention is all you need,'' in
  \emph{Proceedings of Advances in neural information processing systems},
  2017, pp. 5998--6008.

\bibitem{liu2010understanding}
C.~Liu, R.~W. White, and S.~Dumais, ``Understanding web browsing behaviors
  through weibull analysis of dwell time,'' in \emph{Proceedings of the 33rd
  international ACM SIGIR conference on Research and development in information
  retrieval}, 2010, pp. 379--386.

\bibitem{cao2016joint}
K.~Cao and M.~Rei, ``A joint model for word embedding and word morphology,''
  \emph{arXiv preprint arXiv:1606.02601}, 2016.

\bibitem{sennrich2015neural}
R.~Sennrich, B.~Haddow, and A.~Birch, ``Neural machine translation of rare
  words with subword units,'' \emph{arXiv preprint arXiv:1508.07909}, 2015.

\bibitem{bahdanau2014neural}
D.~Bahdanau, K.~Cho, and Y.~Bengio, ``Neural machine translation by jointly
  learning to align and translate,'' \emph{arXiv preprint arXiv:1409.0473},
  2014.

\bibitem{10.1145/3377811.3380327}
\BIBentryALTinterwordspacing
J.~Chen, C.~Chen, Z.~Xing, X.~Xu, L.~Zhu, G.~Li, and J.~Wang, ``Unblind your
  apps: Predicting natural-language labels for mobile gui components by deep
  learning,'' in \emph{Proceedings of the ACM/IEEE 42nd International
  Conference on Software Engineering}, ser. ICSE '20.\hskip 1em plus 0.5em
  minus 0.4em\relax New York, NY, USA: Association for Computing Machinery,
  2020, p. 322–334. [Online]. Available:
  \url{https://doi.org/10.1145/3377811.3380327}
\BIBentrySTDinterwordspacing

\bibitem{liu2018generating}
P.~J. Liu, M.~Saleh, E.~Pot, B.~Goodrich, R.~Sepassi, L.~Kaiser, and
  N.~Shazeer, ``Generating wikipedia by summarizing long sequences,''
  \emph{arXiv preprint arXiv:1801.10198}, 2018.

\bibitem{kitaev2018constituency}
N.~Kitaev and D.~Klein, ``Constituency parsing with a self-attentive encoder,''
  \emph{arXiv preprint arXiv:1805.01052}, 2018.

\bibitem{koehn2004pharaoh}
P.~Koehn, ``Pharaoh: a beam search decoder for phrase-based statistical machine
  translation models,'' in \emph{Proceedings of the Conference of the
  Association for Machine Translation in the Americas}.\hskip 1em plus 0.5em
  minus 0.4em\relax Springer, 2004, pp. 115--124.

\bibitem{tensor2tensor}
A.~Vaswani, S.~Bengio, E.~Brevdo, F.~Chollet, A.~N. Gomez, S.~Gouws, L.~Jones,
  L.~Kaiser, N.~Kalchbrenner, N.~Parmar, R.~Sepassi, N.~Shazeer, and
  J.~Uszkoreit, ``Tensor2tensor for neural machine translation,'' \emph{CoRR},
  vol. abs/1803.07416, 2018.

\bibitem{werbos1990backpropagation}
P.~J. Werbos, ``Backpropagation through time: what it does and how to do it,''
  \emph{Proceedings of the IEEE}, vol.~78, no.~10, pp. 1550--1560, 1990.

\bibitem{kingma2014adam}
D.~P. Kingma and J.~Ba, ``Adam: A method for stochastic optimization,''
  \emph{arXiv preprint arXiv:1412.6980}, 2014.

\bibitem{wang2019learning}
Q.~Wang, B.~Li, T.~Xiao, J.~Zhu, C.~Li, D.~F. Wong, and L.~S. Chao, ``Learning
  deep transformer models for machine translation,'' \emph{arXiv preprint
  arXiv:1906.01787}, 2019.

\bibitem{hochreiter1997long}
S.~Hochreiter and J.~Schmidhuber, ``Long short-term memory,'' \emph{Neural
  computation}, vol.~9, no.~8, pp. 1735--1780, 1997.

\bibitem{ahmad2019context}
W.~U. Ahmad, K.-W. Chang, and H.~Wang, ``Context attentive document ranking and
  query suggestion,'' in \emph{Proceedings of the 42nd International ACM SIGIR
  Conference on Research and Development in Information Retrieval}, 2019, pp.
  385--394.

\bibitem{jiang2018rin}
J.-Y. Jiang and W.~Wang, ``Rin: reformulation inference network for
  context-aware query suggestion,'' in \emph{Proceedings of the 27th ACM
  International Conference on Information and Knowledge Management}, 2018, pp.
  197--206.

\bibitem{chen2018attention}
W.~Chen, F.~Cai, H.~Chen, and M.~de~Rijke, ``Attention-based hierarchical
  neural query suggestion,'' in \emph{The 41st International ACM SIGIR
  Conference on Research \& Development in Information Retrieval}, 2018, pp.
  1093--1096.

\bibitem{ahmad2018multi}
W.~U. Ahmad, K.-W. Chang, and H.~Wang, ``Multi-task learning for document
  ranking and query suggestion,'' in \emph{International Conference on Learning
  Representations}, 2018.

\bibitem{bergstra2011algorithms}
J.~S. Bergstra, R.~Bardenet, Y.~Bengio, and B.~K{\'e}gl, ``Algorithms for
  hyper-parameter optimization,'' in \emph{Advances in neural information
  processing systems}, 2011, pp. 2546--2554.

\bibitem{napoles2015ground}
C.~Napoles, K.~Sakaguchi, M.~Post, and J.~Tetreault, ``Ground truth for
  grammatical error correction metrics,'' in \emph{Proceedings of the 53rd
  Annual Meeting of the Association for Computational Linguistics and the 7th
  International Joint Conference on Natural Language Processing (Volume 2:
  Short Papers)}, 2015, pp. 588--593.

\bibitem{napoles2016gleu}
------, ``Gleu without tuning,'' \emph{arXiv preprint arXiv:1605.02592}, 2016.

\bibitem{papineni2002bleu}
K.~Papineni, S.~Roukos, T.~Ward, and W.-J. Zhu, ``Bleu: a method for automatic
  evaluation of machine translation,'' in \emph{Proceedings of the 40th annual
  meeting on association for computational linguistics}.\hskip 1em plus 0.5em
  minus 0.4em\relax Association for Computational Linguistics, 2002, pp.
  311--318.

\bibitem{dahlmeier2012better}
D.~Dahlmeier and H.~T. Ng, ``Better evaluation for grammatical error
  correction,'' in \emph{Proceedings of the 2012 Conference of the North
  American Chapter of the Association for Computational Linguistics: Human
  Language Technologies}, 2012, pp. 568--572.

\bibitem{oliveto2010equivalence}
R.~Oliveto, M.~Gethers, D.~Poshyvanyk, and A.~De~Lucia, ``On the equivalence of
  information retrieval methods for automated traceability link recovery,'' in
  \emph{Proceedings of 2010 IEEE 18th International Conference on Program
  Comprehension}.\hskip 1em plus 0.5em minus 0.4em\relax IEEE, 2010, pp.
  68--71.

\bibitem{mcmillan2009combining}
C.~McMillan, D.~Poshyvanyk, and M.~Revelle, ``Combining textual and structural
  analysis of software artifacts for traceability link recovery,'' in
  \emph{Proceedings of 2009 ICSE Workshop on Traceability in Emerging Forms of
  Software Engineering}.\hskip 1em plus 0.5em minus 0.4em\relax IEEE, 2009, pp.
  41--48.

\bibitem{gay2009use}
G.~Gay, S.~Haiduc, A.~Marcus, and T.~Menzies, ``On the use of relevance
  feedback in ir-based concept location,'' in \emph{Proceedings of 2009 IEEE
  International Conference on Software Maintenance}.\hskip 1em plus 0.5em minus
  0.4em\relax IEEE, 2009, pp. 351--360.

\bibitem{dit2013integrating}
B.~Dit, M.~Revelle, and D.~Poshyvanyk, ``Integrating information retrieval,
  execution and link analysis algorithms to improve feature location in
  software,'' \emph{Empirical Software Engineering}, vol.~18, no.~2, pp.
  277--309, 2013.

\bibitem{chen2019s}
C.~Chen, Z.~Xing, and Y.~Liu, ``What's spain's paris? mining analogical
  libraries from q\&a discussions,'' \emph{Empirical Software Engineering},
  vol.~24, no.~3, pp. 1155--1194, 2019.

\bibitem{chen2016similartech}
C.~Chen and Z.~Xing, ``Similartech: automatically recommend analogical
  libraries across different programming languages,'' in \emph{Proceedings of
  the 31st IEEE/ACM International Conference on Automated Software
  Engineering}, 2016, pp. 834--839.

\bibitem{chen2019mining}
C.~Chen, Z.~Xing, Y.~Liu, and K.~L.~X. Ong, ``Mining likely analogical apis
  across third-party libraries via large-scale unsupervised api semantics
  embedding,'' \emph{IEEE Transactions on Software Engineering}, 2019.

\bibitem{chen2020similarapi}
C.~Chen, ``Similarapi: mining analogical apis for library migration,'' in
  \emph{2020 IEEE/ACM 42nd International Conference on Software Engineering:
  Companion Proceedings (ICSE-Companion)}.\hskip 1em plus 0.5em minus
  0.4em\relax IEEE, 2020, pp. 37--40.

\bibitem{chen2020wireframe}
J.~Chen, C.~Chen, Z.~Xing, X.~Xia, L.~Zhu, J.~Grundy, and J.~Wang,
  ``Wireframe-based ui design search through image autoencoder,'' \emph{ACM
  Transactions on Software Engineering and Methodology (TOSEM)}, vol.~29,
  no.~3, pp. 1--31, 2020.

\bibitem{chen2019gallery}
C.~Chen, S.~Feng, Z.~Xing, L.~Liu, S.~Zhao, and J.~Wang, ``Gallery dc: Design
  search and knowledge discovery through auto-created gui component gallery,''
  \emph{Proceedings of the ACM on Human-Computer Interaction}, vol.~3, no.
  CSCW, pp. 1--22, 2019.

\bibitem{chen2020lost}
C.~Chen, S.~Feng, Z.~Liu, Z.~Xing, and S.~Zhao, ``From lost to found: Discover
  missing ui design semantics through recovering missing tags,''
  \emph{Proceedings of the ACM on Human-Computer Interaction}, vol.~4, no.
  CSCW2, pp. 1--22, 2020.

\bibitem{nie2016query}
L.~Nie, H.~Jiang, Z.~Ren, Z.~Sun, and X.~Li, ``Query expansion based on crowd
  knowledge for code search,'' \emph{IEEE Transactions on Services Computing},
  vol.~9, no.~5, pp. 771--783, 2016.

\bibitem{lu2015query}
M.~Lu, X.~Sun, S.~Wang, D.~Lo, and Y.~Duan, ``Query expansion via wordnet for
  effective code search,'' in \emph{Proceedings of 2015 IEEE 22nd International
  Conference on Software Analysis, Evolution, and Reengineering}.\hskip 1em
  plus 0.5em minus 0.4em\relax IEEE, 2015, pp. 545--549.

\bibitem{rahman2019automatic}
M.~M. Rahman, C.~K. Roy, and D.~Lo, ``Automatic query reformulation for code
  search using crowdsourced knowledge,'' \emph{Empirical Software Engineering},
  vol.~24, no.~4, pp. 1869--1924, 2019.

\bibitem{rahman2018effective}
M.~M. Rahman and C.~Roy, ``Effective reformulation of query for code search
  using crowdsourced knowledge and extra-large data analytics,'' in
  \emph{Proceedings of 2018 IEEE International Conference on Software
  Maintenance and Evolution}.\hskip 1em plus 0.5em minus 0.4em\relax IEEE,
  2018, pp. 473--484.

\bibitem{haiduc2011effect}
S.~Haiduc and A.~Marcus, ``On the effect of the query in ir-based concept
  location,'' in \emph{Proceedings of 2011 IEEE 19th International Conference
  on Program Comprehension}.\hskip 1em plus 0.5em minus 0.4em\relax IEEE, 2011,
  pp. 234--237.

\bibitem{haiduc2012evaluating}
S.~Haiduc, G.~Bavota, R.~Oliveto, A.~Marcus, and A.~De~Lucia, ``Evaluating the
  specificity of text retrieval queries to support software engineering
  tasks,'' in \emph{Proceedings of 2012 34th International Conference on
  Software Engineering}.\hskip 1em plus 0.5em minus 0.4em\relax IEEE, 2012, pp.
  1273--1276.

\bibitem{haiduc2012automatic}
S.~Haiduc, G.~Bavota, R.~Oliveto, A.~De~Lucia, and A.~Marcus, ``Automatic query
  performance assessment during the retrieval of software artifacts,'' in
  \emph{Proceedings of the 27th IEEE/ACM international conference on Automated
  Software Engineering}, 2012, pp. 90--99.

\bibitem{haiduc2013automatic}
S.~Haiduc, G.~Bavota, A.~Marcus, R.~Oliveto, A.~De~Lucia, and T.~Menzies,
  ``Automatic query reformulations for text retrieval in software
  engineering,'' in \emph{Proceedings of 2013 35th International Conference on
  Software Engineering}.\hskip 1em plus 0.5em minus 0.4em\relax IEEE, 2013, pp.
  842--851.

\bibitem{haiduc2013query}
S.~Haiduc, G.~De~Rosa, G.~Bavota, R.~Oliveto, A.~De~Lucia, and A.~Marcus,
  ``Query quality prediction and reformulation for source code search: The
  refoqus tool,'' in \emph{Proceedings of 2013 35th International Conference on
  Software Engineering}.\hskip 1em plus 0.5em minus 0.4em\relax IEEE, 2013, pp.
  1307--1310.

\bibitem{rahman2016quickar}
M.~M. Rahman and C.~K. Roy, ``Quickar: automatic query reformulation for
  concept location using crowdsourced knowledge,'' in \emph{Proceedings of 2016
  31st IEEE/ACM International Conference on Automated Software
  Engineering}.\hskip 1em plus 0.5em minus 0.4em\relax IEEE, 2016, pp.
  220--225.

\bibitem{li2016query}
Z.~Li, T.~Wang, Y.~Zhang, Y.~Zhan, and G.~Yin, ``Query reformulation by
  leveraging crowd wisdom for scenario-based software search,'' in
  \emph{Proceedings of the 8th Asia-Pacific Symposium on Internetware}, 2016,
  pp. 36--44.

\bibitem{chen2016learning}
G.~Chen, C.~Chen, Z.~Xing, and B.~Xu, ``Learning a dual-language vector space
  for domain-specific cross-lingual question retrieval,'' in \emph{2016 31st
  IEEE/ACM International Conference on Automated Software Engineering
  (ASE)}.\hskip 1em plus 0.5em minus 0.4em\relax IEEE, 2016, pp. 744--755.

\bibitem{howard2013automatically}
M.~J. Howard, S.~Gupta, L.~Pollock, and K.~Vijay-Shanker, ``Automatically
  mining software-based, semantically-similar words from comment-code
  mappings,'' in \emph{Proceedings of 2013 10th Working Conference on Mining
  Software Repositories}.\hskip 1em plus 0.5em minus 0.4em\relax IEEE, 2013,
  pp. 377--386.

\bibitem{rahman2017improved}
M.~M. Rahman and C.~K. Roy, ``Improved query reformulation for concept location
  using coderank and document structures,'' in \emph{Proceedings of 2017 32nd
  IEEE/ACM International Conference on Automated Software Engineering}.\hskip
  1em plus 0.5em minus 0.4em\relax IEEE, 2017, pp. 428--439.

\bibitem{sirres2018augmenting}
R.~Sirres, T.~F. Bissyand{\'e}, D.~Kim, D.~Lo, J.~Klein, K.~Kim, and
  Y.~Le~Traon, ``Augmenting and structuring user queries to support efficient
  free-form code search,'' \emph{Empirical Software Engineering}, vol.~23,
  no.~5, pp. 2622--2654, 2018.

\end{thebibliography}

\end{document}